\documentclass[sigconf]{acmart}
\AtBeginDocument{%
  }

\setcopyright{rightsretained}
\copyrightyear{2025}
\acmYear{2025}
\acmDOI{XXXXXXX.XXXXXXX}
\acmConference[ASIA CCS '26]{}{June 1--6, 2026}{Bangalore, IN}
\acmISBN{978-1-4503-XXXX-X/2018/06}

\usepackage{amsmath,amsfonts}
\usepackage{algorithmic}
\usepackage{graphicx}
\usepackage{textcomp}
\usepackage{xcolor}
\usepackage{acronym}
\usepackage{subfig}
\usepackage{pifont}
\usepackage{balance}

\acrodef{RFF}{Radio Frequency Fingerprinting}
\acrodef{DF}{Device Fingerprint}
\acrodef{RF}{Radio Frequency}
\acrodef{ML}{Machine Learning} 
\acrodef{DL}{Deep Learning}
\acrodef{BPSK}{Binary Phase Shift Keying}
\acrodef{SDR}{Software Defined Radio}
\acrodef{SNR}{Signal to Noise Ratio}
\acrodef{CNN}{Convolutional Neural Network}
\acrodef{AE}{Autoencoders}
\acrodef{FPR}{False Positive Ratio}
\acrodef{FNR}{False Negative Ratio}
\acrodef{MSE}{Mean Square Error}
\acrodef{IQ}{In-Phase - Quadrature}
\acrodef{CFO}{Carrier Frequency Offset}
\acrodef{AML}{Adversarial Machine Learning}
\acrodef{CSI}{Channel State Information}
\acrodef{LSTM}{Long Short-Term Memory}
\acrodef{ROC}{Receiver Operating Characteristic}
\acrodef{CFO}{Carrier Frequency Offset}

\newcommand{\sol}{HidePrint}
\newcommand{\xmark}{\ding{55}}

\begin{document}

\title{HidePrint: Protecting Device Anonymity by Obscuring Radio Fingerprints}

\author{Gabriele Oligeri}
\orcid{}
\affiliation{%
  \institution{Hamad Bin Khalifa University}
  \city{Doha}
  \country{Qatar}}
\email{goligeri@hbku.edu.qa}

\author{Savio Sciancalepore}
\orcid{}
\affiliation{%
  \institution{Eindhoven University of Technology}
  \city{Eindhoven}
  \country{Netherlands}
}
\email{s.sciancalepore@tue.nl}



\begin{abstract}
Radio Frequency Fingerprinting (RFF) techniques allow a receiver to authenticate a transmitter by analyzing the physical layer of the radio spectrum. Although the vast majority of scientific contributions focus on improving the performance of RFF considering different parameters and scenarios, in this work, we consider RFF as an attack vector to identify a target device in the radio spectrum. \\
We propose, implement, and evaluate {\em HidePrint}, a solution to prevent identification through RFF without affecting the quality of the communication link between the transmitter and the receiver. {\em HidePrint} hides the transmitter's fingerprint against an illegitimate eavesdropper through the injection of controlled noise into the transmitted signal. 
We evaluate our solution against various state-of-the-art RFF techniques, considering several adversarial models, data from real-world communication links (wired and wireless), and protocol configurations. Our results show that the injection of a Gaussian noise pattern with a normalized standard deviation of (at least) 0.02 prevents device fingerprinting in all the considered scenarios, while affecting the Signal-to-Noise Ratio (SNR) of the received signal by only 0.1 dB. Moreover, we introduce {\em selective radio fingerprint disclosure}, a new technique that allows the transmitter to disclose the radio fingerprint to only a subset of intended receivers. 
\end{abstract}

\maketitle

\section{Introduction}
\label{sec:introduction}

\ac{RFF} techniques are today a popular tool used to differentiate and identify radio devices through the unique characteristics inherent in the \ac{RF} emissions~\cite{jagannath2022_comnet}. These unique characteristics come from the hardware of the device, and in particular, from the small differences featured by the hardware components during the manufacturing process. Such differences lead to the main idea that two electronic devices that are exactly alike do not exist because of physical differences in oscillators, amplifiers, and built-in antennas. Since these differences are reflected in the over-the-air signal, a receiver can extract them through the analysis of the received signals (\ac{IQ} samples) and leverage them to uniquely identify the transmitter~\cite{alhazbi2024_iwcmc}.

A strict requirement for all \ac{RFF} systems is the ability to collect information from the physical layer of the communication link~\cite{xu2016_comst}. Such features persist on the receiver side up to the conversion from symbols to bits, making the availability of such information a requirement for \ac{RFF}. At the physical layer, the baseband signal can be expressed via \ac{IQ} samples, representing the in-phase (I) and quadrature (Q) components of the signal, used to carry the information associated with the transmitted bit string and some additional information such as noise, attenuation, Doppler shift, and finally, the radio fingerprint~\cite{shen2022_tifs}.

\ac{RFF} techniques are characterized by several key advantages compared to traditional cryptographic solutions. RFF is crypto-less, i.e., it requires no implementation of cryptographic techniques, and, most importantly, it requires no key distribution, thus solving by design the challenges related to joining and leaving entities from the network. Moreover, \ac{RFF} does not require any intervention on the transmitter side. The transmitter is authenticated through the information provided by its transmitted signal. In contrast, \ac{RFF} requires a receiver capable of collecting information from the physical layer and a \ac{DL} model. The receiver should train a classifier with legitimate signals (\ac{IQ} samples) in order to challenge it in the subsequent phase, with a test set (of \ac{IQ} samples) to assess either the presence or the absence of the transmitter(s)~\cite{danev2012_csur}.

Recent scientific literature mostly focused on maximizing the performance of \ac{RFF} techniques while addressing several real-world challenges, e.g., robustness to multipath fading~\cite{restuccia2019_manet}, jamming~\cite{smailes2024_arxiv}, time-varying distortions~\cite{h2023_twc}, reliability over time~\cite{alhazbi2023_acsac}, and mobility of the transmitter~\cite{oligeri2023tifs}, to name a few. As a result, today, \ac{RFF} solutions are becoming increasingly effective for successfully classifying and authenticating the transmitter. This makes \ac{RFF} a promising candidate to implement authentication at the physical layer while mitigating the common challenges of key distribution in dense networks and implementation in resource-constrained devices.
While a ready-to-market \ac{RFF} solution is not yet available, the accuracy of the proposed techniques is increasing for the most diverse scenarios and applications~\cite{alhazbi2024_iwcmc}. 

In this work, we take a different perspective on \ac{RFF}, exploring the possibility that a malicious party could use \ac{RFF} to detect and track the presence of a device in the radio spectrum. Indeed, an adversary could train a model on signals generated by a specific (pool of) device(s) and then leverage such a model to detect the presence of the transmitter in the radio spectrum---making useless any anonymization technique implemented by higher communications layers, e.g., MAC address anonymization.
Malicious detection and tracking of devices pose significant challenges to people privacy, enabling mass and targeted surveillance. For example, adversaries could track when and where a device appears in the radio spectrum to infer a person's daily routine, places of interest, and social interactions~\cite{baig2025_tai},~\cite{ardoin2025_arxiv}. 

Research on protection against \ac{RFF}-based attacks is relatively scarce in the literature and is mostly qualitative. Pioneering contributions consider \ac{RFF} techniques for WiFi based on \ac{CSI} fingerprinting~\cite{danev2010_wisec,abanto2020_macs}, while more recent work explore the theoretical usage of \ac{AML} techniques to generate perturbations in the transmitted signals thwarting \ac{RFF} while keeping the communication quality acceptable~\cite{papangelo2023commag},~\cite{liu2023_wcl},~\cite{sun2022_rs},~\cite{lu2024_arxiv}. However, such works either focus on pure data analysis without considering the technical challenges of data manipulation and wireless transmission 
or use simplistic \ac{RFF} models. As a result, no contributions have investigated the potential of intentionally altering the radio fingerprint to prevent the tracking of a general-purpose transmitter while allowing authorized receivers to authenticate it without affecting communication quality.

{\bf Research Questions.} In this work, we formulate the problem mentioned above through two main research questions. 

{\em RQ1: Can we prevent a malicious adversary from successfully using \ac{RFF} to identify and track a (pool of) device(s)?}

{\em RQ2: Can we enable selective radio fingerprint disclosure?}

Since \ac{RFF} requires no shared secrets, the legitimate and malicious receivers leverage the same information. Thus, a solution preventing \ac{RFF}-based attacks by an adversary (RQ1) affects the ability of a legitimate device to authenticate the transmitter(s), thus yielding to our second research question (RQ2). Selective radio fingerprint disclosure could allow a designated set of (legitimate) receivers to perform \ac{RFF} while excluding other (malicious) ones.

{\bf Contribution.} This work provides manifold contributions:
\begin{itemize}
    \item We design, implement, evaluate, and test two \ac{RFF}-based attacks in different scenarios, considering both an (ideal) wired and a (real) wireless communication link. The wired setup allows us to capture the precise impact of our attacks and study in a controlled setup, while the wireless setup is used to confirm such findings in the real-world scenario.
    \item We consider two main adversary models as a function of the ability of the adversary to collect data from either a target device or the complete pool of devices.
    \item We consider an adversary collecting data for \ac{RFF} both before and after anonymization, so updating at runtime the \ac{RFF} models to de-anonymize the transmitters.
    \item We provide the theoretical framework, description, real implementation, and evaluation of {\em \sol}, a solution to prevent unauthorized device fingerprinting.
    \item Inspired by \sol, we propose a technique enabling selective radio fingerprint disclosure, thus allowing only a selected subset of legitimate receivers to perform \ac{RFF}. 
\end{itemize}

{\bf Paper organization.} The paper is organized as follows. Sect.~\ref{sec:related_work} reviews related work, Sect.~\ref{sec:adversarial_model} introduces the scenario and adversary models, Sect.~\ref{sec:background} introduces the rationale of our solution, Sect.~\ref{sec:measurement_setup} describes our measurement campaign, Sect.~\ref{sec:device_fingerprinting} describes our methodology, Sect.~\ref{sec:results} discusses the performance of our solution, Sect.~\ref{sec:comparison} compares our solution to competing approaches, Sect.~\ref{sec:hideprint} presents selective radio fingerprint disclosure, Sect~\ref{sec:discussion} discusses the main results and limitations, and finally, Sect.~\ref{sec:conclusion} concludes the paper. 

\section{Related Work}
\label{sec:related_work}

\begin{table*}
\footnotesize
\centering
\begin{tabular}{c|c|c|c|c|c|c|c}
\multicolumn{1}{c|}{\textbf{Ref.}} & \multicolumn{1}{c|}{\textbf{\begin{tabular}[c]{@{}c@{}}Protection \\ against RFF\end{tabular}}} & \multicolumn{1}{c|}{\textbf{\begin{tabular}[c]{@{}c@{}}RFF \\ Technique\end{tabular}}} & \multicolumn{1}{c|}{\textbf{\begin{tabular}[c]{@{}c@{}}Protection \\ Strategy\end{tabular}}} & \multicolumn{1}{c|}{\textbf{\begin{tabular}[c]{@{}c@{}} No External \\  Devices\end{tabular}}} & \multicolumn{1}{c|}{\textbf{\begin{tabular}[c]{@{}c@{}}Evaluation \\ Strategy\end{tabular}}} & \multicolumn{1}{c|}{\textbf{\begin{tabular}[c]{@{}c@{}}Selective \\ Radio Fingerprint Disclosure\end{tabular}}} & \multicolumn{1}{c}{\textbf{\begin{tabular}[c]{@{}c@{}}Technology\\ Independent\end{tabular}}} \\ \hline
\cite{danev2010_wisec} & \xmark & CSI-based & Phase Correction & \xmark & Real-world tests & \xmark & \xmark \\ \hline
\cite{abanto2020_macs} &  \checkmark & CSI-based & Random Phase Correction & \checkmark & Real-world tests & \checkmark & \xmark  \\ \hline
\cite{papangelo2023commag} & \xmark  & Image-based &  \ac{AML}  &  \checkmark    &  Simulations   &  \xmark &  \checkmark  \\ \hline
\cite{liu2023_wcl} &  \xmark  & Raw \ac{IQ} & \ac{AML} & \checkmark  & Simulations &  \xmark &  \checkmark \\ \hline
\cite{sun2022_rs} &  \xmark  &  Raw \ac{IQ} & \ac{AML} & \checkmark & Simulations &  \xmark  &  \checkmark \\ \hline
\cite{lu2024_arxiv} &  \checkmark   &  Raw \ac{IQ}   &  \ac{AML} &  \checkmark  & Real-world tests  & \xmark &  \checkmark \\ \hline
\cite{givehchian2024_sp} & \checkmark & CFO & CFO Randomization & \checkmark & Real-world tests & \xmark & \xmark \\
\hline
\cite{irfan2024_asiaccs} & \checkmark   &  Image-based  & Friendly Jamming & \xmark  & Real-world tests  & \xmark  &  \checkmark \\ \hline
\begin{tabular}[c]{@{}l@{}} {\bf This}\\ {\bf Paper}\end{tabular} & \checkmark & Image-based & Random Noise  &  \checkmark & Real-world tests & \checkmark & \checkmark
\end{tabular}
\caption{Qualitative comparison of scientific contributions focusing on RFF disruption.}
\label{tab:related}
\end{table*}

Early research on RFF investigated the usage of specific radio signal imperfections to identify transmitters and reject spoofing attacks, e.g., \ac{IQ} imbalance, phase shift, and \ac{CFO}, mostly using \ac{ML} models~\cite{reising2015_tifs},~\cite{zhuang2018_asiaccs},~\cite{reising2020_iotj}.
With the diffusion of \ac{DL}, research on \ac{RFF} shifted towards using raw physical-layer signals, i.e., \ac{IQ} samples, due to the enhanced capability of such algorithms to extract hidden patterns~\cite{jagannath2022_comnet}. Many solutions leverage raw \ac{IQ} samples directly as input to various \ac{DL} classifiers, e.g., \acp{CNN}~\cite{alshawabka2020_infocom}~\cite{solenthaler2025_spacesec} and \acp{AE}~\cite{smailes2023_ccs}. On the one hand, such solutions are characterized by outstanding classification and authentication performance, reaching accuracy values very close to perfect discrimination~\cite{xu2016_comst}. On the other hand, recent research showed that these models are very unstable, being significantly affected by any source of noise, introduced, e.g., by the warm-up of the radio~\cite{elmaghbub2024_wisec}, hardware reset~\cite{alhazbi2023_acsac}, temperature changes in the environment~\cite{bechir_2024_tmlcn}~\cite{gu2024_tosn}, and fluctuations in channel conditions. In this context, image-based \ac{RFF} techniques feature enhanced flexibility and portability across different real-world scenarios and configurations. Recent literature has shown that pre-processing (and converting) raw \ac{IQ} samples into images leads to the creation of \ac{DL}-based models robust to \ac{AML}~\cite{papangelo2023commag}, restart of the radios~\cite{alhazbi2023_acsac}, and firmware reload~\cite{irfan2024_arxiv_reliability}, to name a few. Image-based \ac{RFF} techniques have also been used beyond device identification, e.g., for channel fingerprinting~\cite{oligeri2024sac, sadighian2024ccnc} and jamming detection~\cite{alhazbi2023ccnc,sciancalepore2023jamming}.

Only a few works in the literature considered \ac{RFF} as an attack and, thus, investigated how to protect a device from unconditionally disclosing the radio fingerprint to malicious parties. An early attempt in this direction is the contribution by Abanto et al.~\cite{abanto2020_macs}, showing that the injection of random phase corrections into the signal emitted by WiFi devices can reduce the accuracy of \ac{RFF} systems using \ac{CSI} to authenticate WiFi devices. The \ac{RFF} algorithm considered in their paper only considers shift estimation for WiFi devices, and DL-based solutions are not considered. Also, the pioneering contribution by Danev et al. in~\cite{danev2010_wisec} considers CSI-based \ac{RFF} and experiments with impersonation attacks via powerful signal generators. More recently, contributions such as the ones by Papangelo et al.~\cite{papangelo2023commag}, Sun et al.~\cite{sun2022_rs}, and Liu et al.~\cite{liu2023_wcl} investigated the effectiveness of \ac{AML} to disrupt the correct operation of \ac{RFF}. The only recent work considering the usage of adversarial techniques to conceal radio fingerprints is the preliminary work by Lu et al. in~\cite{lu2024_arxiv}, providing an early solution to anonymization against \ac{RFF}. The authors propose to generate a perturbation signal $\delta$ calculated according to a predefined strategy that matches the decision algorithm taken by the classifier. The adopted \ac{CNN} network, designed from scratch, uses raw \ac{IQ} samples, sharing the same limitations described above. Moreover, the authors use a limited set of devices (five) of different brands, easing the design of the RFF model and the following anonymization task, and use higher-order modulation schemes, increasingly sensitive to even low-power noise. Recently, Givehchian et al.~\cite{givehchian2024_sp} demonstrated that RFF methods based on the \ac{CFO} can be thwarted by randomizing the \ac{CFO} of Bluetooth and WiFi devices. However, their solution only works against RFF via \ac{CFO}. An alternative approach is the proposal by Irfan et al.~\cite{irfan2024_asiaccs}, using an external device (jammer) to protect against \ac{RFF}. Although effective, such an approach requires an external device, complicating the solution design. Thus, as summarized in Tab.~\ref{tab:related}, no work in the literature investigated how to protect from malicious \ac{RFF} while considering various adversary models, characterized by increasing knowledge of the \ac{RFF} model of the target device(s), and state-of-the-art image-based RFF techniques, without using additional devices. Moreover, there is currently no technique providing selective radio fingerprint disclosure at the physical layer without exploiting information coming from the higher layers, e.g., channel state information~\cite{abanto2020_macs}, so being  independent of the adopted higher-layer technology. We address this gap in the remainder of this paper.

\section{Scenario and Adversarial model}
\label{sec:adversarial_model}
\begin{figure}[t]
    \centering
    \includegraphics[width=\columnwidth, angle = 0,trim = 0mm 0mm 0mm 0mm]{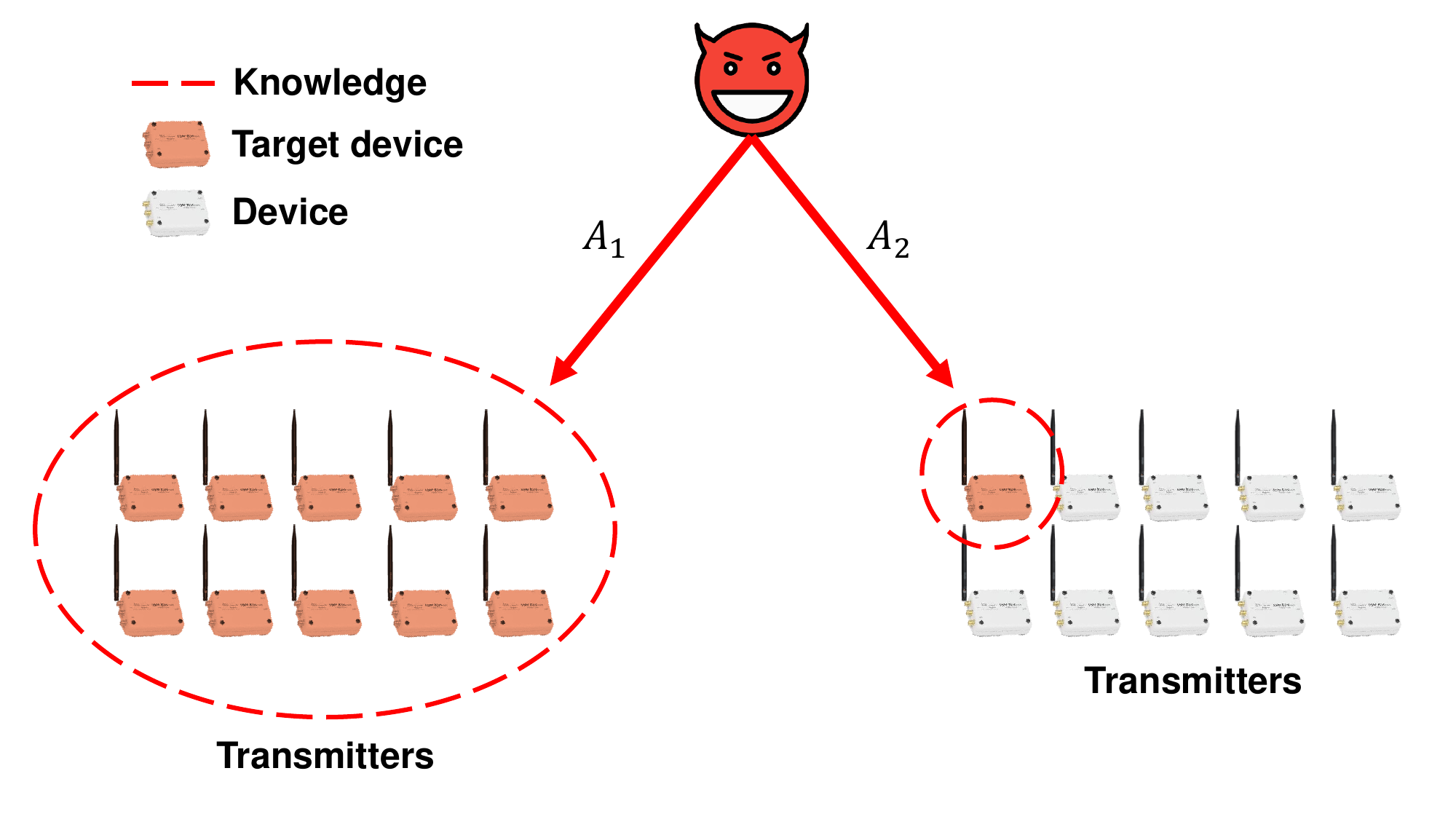}
    \caption{Adversary model. We consider two realistic assumptions: (i) the adversary owns a model containing features from all the devices in the pool ($\mathcal{A}_1$) and wants to de-anonymize any device, and (ii) the adversary owns a model containing features from one only device ($\mathcal{A}_2$) and wants to detect its presence in the radio spectrum.} 
    \label{fig:adversary}
\end{figure}

Our scenario is constituted by a pool of ten (10) transmitters that aim to remain anonymous while communicating. Current state-of-the-art techniques already allow enforcing anonymity at higher layers of the ISO OSI communication stack, e.g., at the network layer via VPN and at the MAC layer via MAC randomization, while our scenario specifically considers anonymity against physical layer identification. Without loss of generality, we consider the use of the \ac{BPSK} modulation scheme. Many modern communication technologies currently use \ac{BPSK}, including IEEE 802.15.4 (Zigbee, Thread), IEEE 802.11 (WiFi), and many satellite communication systems (e.g., IRIDIUM) due to its robustness to noise introduced by the wireless channel~\cite{rappaport2024_book}. Indeed, \ac{BPSK} is able to handle the highest noise level (or distortions) before the demodulator takes the wrong decision on the symbol, thus representing the best modulation candidate for our objective. A transmitter willing to achieve anonymity (at the physical layer) could also switch to \ac{BPSK}, thus trading off bandwidth (lower communication rate) with privacy.
Moreover, we consider an adversary able to deploy state-of-the-art image-based \ac{RFF} techniques to identify and (potentially) track devices in the radio spectrum. We recall that, as discussed in Sec.~\ref{sec:related_work}, image-based \ac{RFF} models typically outperform other solutions thanks to their enhanced robustness to noise and flexibility to real-world phenomena. Figure~\ref{fig:adversary} shows our adversary model based on two different assumptions related to its a priori knowledge.

{\bf Adversary 1 ($\mathcal{A}_1$).} The adversary collects data from all devices in the pool of the transmitters, with the objective of identifying their presence in the communication link, thus deanonymizing them. This objective can be formulated as a classical multiclass classification problem, where the adversary collects physical layer information from the 10 transmitters in the pool, trains a model with 10 classes (one for each transmitter), and then aims at maximizing its chances to deanonymize the transmitters (i.e., classifying them with an accuracy higher than the random guess, equal to 0.1).

{\bf Adversary 2 ($\mathcal{A}_2$).} The adversary focuses the attack on a target device, thus collecting the data related to a specific transmitter from the physical layer of the communication link. Such data is used for training a \ac{DL} model to be used to detect the presence of such a specific transmitter in the wild. In line with the current literature~\cite{oligeri2023tifs}~\cite{smailes2023_ccs}, we address this challenge as a one-class classification problem, thus considering \emph{autoencoders} as our identification tool.

\section{Rationale of our Solution}
\label{sec:background}
\begin{figure}[t]
    \centering
    \includegraphics[width=\columnwidth, angle = 0,trim = 30mm 85mm 30mm 85mm]{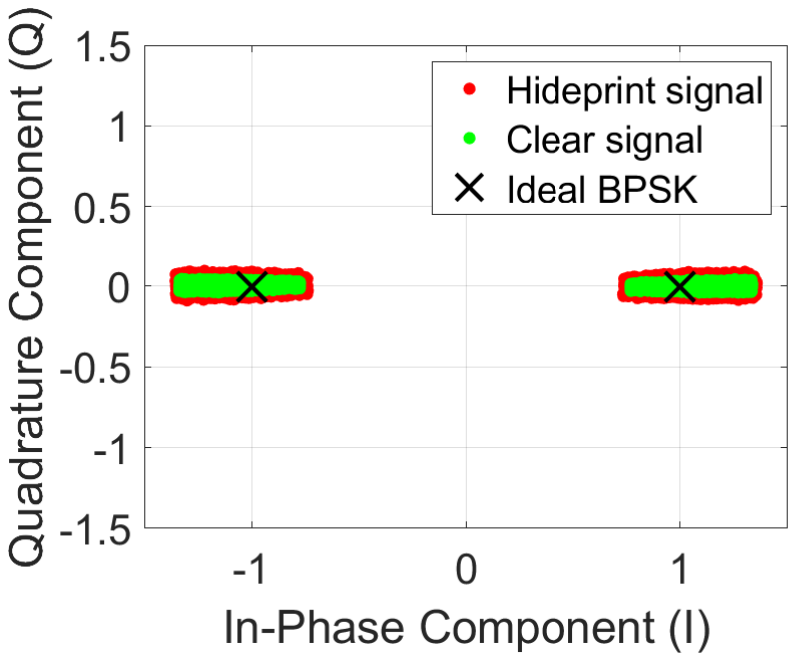}
    \caption{A toy example of our \sol\ solution: the device fingerprint is removed from the clear signal (green area) by adding random noise, thus obtaining the signal dispersed in the red area (signal reshaped by \sol).} 
    \label{fig:solutionIQ}
\end{figure}

Let $I(t)$ and $Q(t)$ be the in-phase and quadrature components (\ac{IQ} components) of the baseband signal. The associated passband transmitted signal can be expressed as per Eq.~\ref{eq:band_pass_signal_IQ}:
\begin{equation}
    x(t) = I(t) \cos{(\omega_c t)} + Q(t) \sin{(\omega_c t)},
    \label{eq:band_pass_signal_IQ}
\end{equation}
where $\omega_c = 2\pi f_c$ and $f_c$ is the signal carrier. Equation~\ref{eq:band_pass_signal_IQ} can be rewritten as Eq.~\ref{eq:band_pass_signal}:
\begin{equation}
    x(t) = \alpha(t) \cos{(\omega_c t + \phi(t))},
    \label{eq:band_pass_signal}
\end{equation}
where $\alpha(t) = \sqrt{I(t)^2 + Q(t)^2}$ and $\phi(t) = \tan^{-1}{\frac{Q(t)}{I(t)}}$ are the amplitude and phase associated with the in-phase ($I(t)$) and quadrature ($Q(t)$) components of the baseband signal, respectively.
We consider the \ac{BPSK} modulation scheme, and therefore, we can recover the baseband signal $r(t)$ from $x(t)$ as:
\begin{eqnarray}
    r(t) & = & x(t) \cdot \cos{(\omega_c t)} \nonumber\\
         & = & \alpha(t) \cos{(\omega_c t + \phi(t))} \cos{(\omega_c t)} \nonumber\\
         & = & \frac{1}{2} \alpha(t) \cos{(\phi(t))}, \label{eq:received_signal}
\end{eqnarray}
where we applied low-pass filtering to remove the component at $f = 2f_c$. We now assume that $\alpha(t)$ and $\phi(t)$ are affected by synthetically-generated noise with components $n_\alpha(t)$ and $n_\phi(t)$, i.e.:
\begin{eqnarray}
    \alpha(t) & = & \alpha'(t) + n_\alpha(t), \nonumber \\
    \phi(t) & = & \phi'(t) + n_\phi(t), \label{eq:fingerprint_components} 
\end{eqnarray}
where $\alpha'(t)$ and $\phi'(t)$ represent the actual amplitude and phase components of the baseband signal, respectively. We recall that $Q(t) = 0$, with \ac{BPSK}. Under ideal conditions, Eq.~\ref{eq:fingerprint_components} becomes:
\begin{eqnarray}
    \alpha(t) & = & 1 + \epsilon_\alpha(t) + n_\alpha(t), \nonumber \\
    \phi(t) & = & \{0, \pi\} + \epsilon_\phi(t) + n_\phi(t), \label{eq:fingerprint_components_ideal}
\end{eqnarray}
where we considered the amplitude component $\alpha'(t) = 1 + \epsilon_\alpha(t)$, i.e., a phasor (unitary amplitude) affected by fingerprint $\epsilon_\alpha(t)$ and phase $\phi'(t) = \{0, \pi\} + \epsilon_\phi(t)$, taking on the values either $0$ or $\pi$ as a function of the transmitted bit, while $\epsilon_\phi(t)$ is the phase component. 
Note that $n_\alpha(t)$ and $n_\phi(t)$ directly affect the components of the fingerprint $\epsilon_\alpha(t)$ and $\epsilon_\phi(t)$. Therefore, the transmitter could hide such components by synthetically injecting noise into the baseband signal. Finally, we highlight that our analysis does not consider any noise coming from the channel, as well as multipath. We prove the reliability of our solution in the later sections of this paper, using real wireless measurements.
As a toy example, consider Fig.~\ref{fig:solutionIQ}, where we compare: (i) the position of the \ac{IQ} samples for the ideal BPSK modulation scheme (black crosses), (ii) an actual clear signal $r(t)$ (green area), and finally, (iii) the noise-obfuscated signal (red area), where the original signal $r(t)$ has been subjected to the noise components $n_\alpha(t)$ and $n_\phi(t)$. We highlight that the figure is the result of real communication in a wired link, where we applied a Gaussian noise of standard deviation 0.05 simultaneously on the real and imaginary parts of the signal ($10^5$ \ac{IQ} samples). In the remainder of this work, we show that such noise intensity is enough to remove the radio fingerprint of the transmitter while negligibly affecting the quality of the signal, i.e., the \ac{SNR}.

\section{Measurement setup}
\label{sec:measurement_setup}

\begin{figure}[t]
    \centering
    \includegraphics[width=\columnwidth, angle = 0,trim = 10mm 0mm 10mm 10mm]{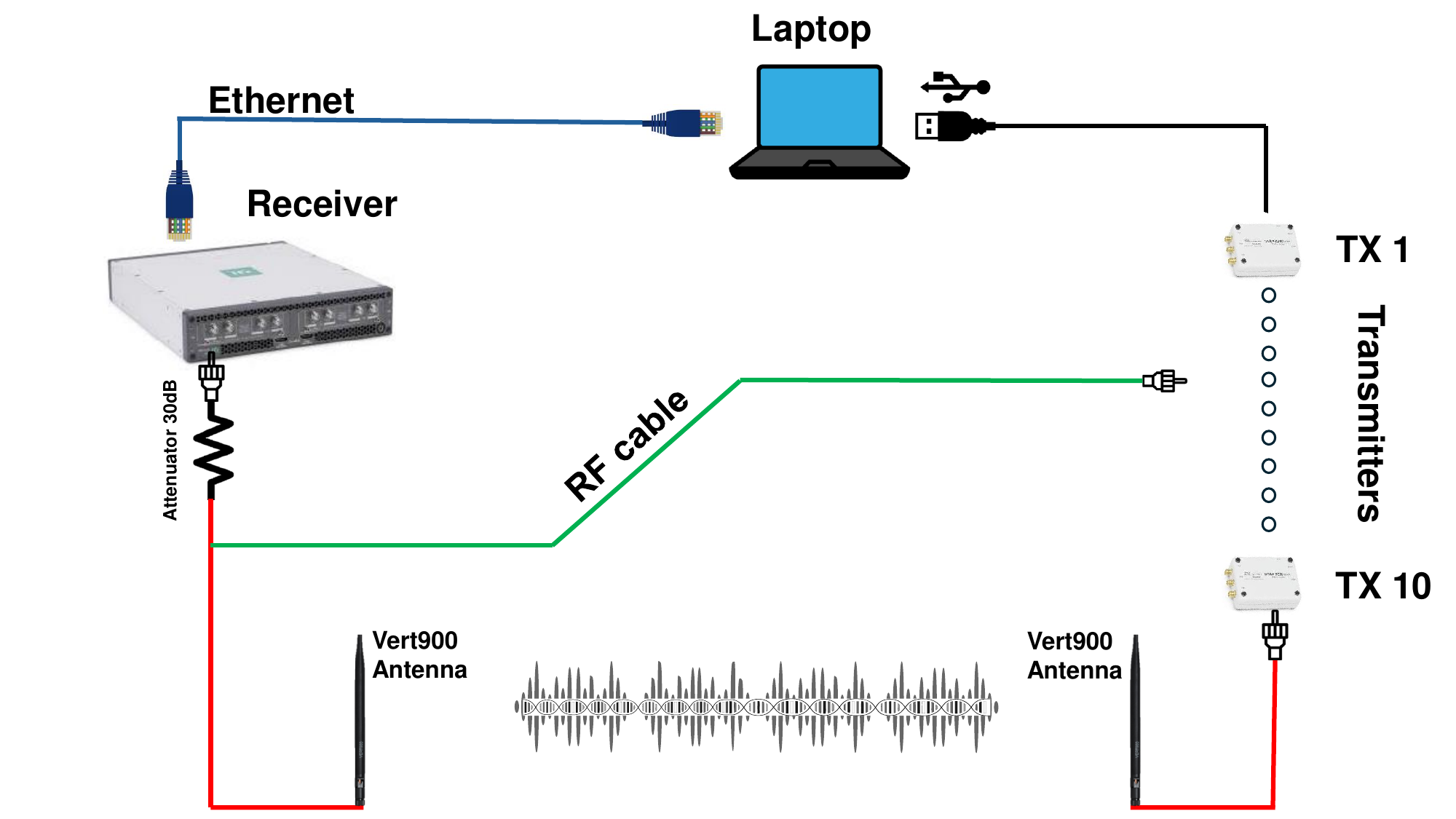}
    \caption{We use 10 transmitters (USRP B200-mini-i) connected (one per time) to the receiver (USRP X410). We consider two scenarios: a wired connection between the transmitter and the receiver (green) and a wireless link with two Vert900 antennas (red). For both, we use 30dB attenuation.} 
    \label{fig:cable_scenario}
\end{figure}

{\bf Software and hardware setup.} Our measurement setup consists of a receiver Ettus USRP X410~\cite{X410_datasheet}, and ten (10) transmitters USRP B200-mini-i~\cite{b200_datasheet} (see Fig.~\ref{fig:cable_scenario}). We consider a general-purpose laptop with Ubuntu 24.04 LTS running GNURadio 3.10 (with Python 3.12) and the related flowcharts to control the receiver and each of the transmitters (one per time).
The flowchart for the transmitter (USRP B200-mini-i) involves a standard \ac{BPSK} modulation scheme, transmitting a repeating sequence of bytes spanning between 0 and 255, generated through the block {\em Vector Source}. The sequence is processed and converted to a stream of bits and, finally, to a stream of complex numbers where the noise $(n_\alpha(t), n_\phi(t))$ (recall Sect.~\ref{sec:background}) is added before the {\em Root Raised Cosine Filter}. Finally, we use a {\em USRP Sink} block to transmit the signal. The RRC filter performs an upsampling of 4 samples per symbol, i.e., from a symbol rate of 250K \ac{IQ} symbols per second to $10^6$ samples per second. All our measurements consider a carrier frequency $f_c = 900$~MHz and a normalized transmitter power of 0.7.
The flowchart for the receiver involves a {\em USRP Source} block with sample rate $10^6$ (same as the transmitter, according to the modified Nyquist theorem for complex signals~\cite{sampling_same}), carrier frequency $f_c = 900$~MHz, and normalized receiver gain of 0.7, followed by a {\em Costas Loop} block, an adaptive gain control {\em AGC} block, a {\em Root Raised Cosine Filter} downsampling from 1~M samples per second (sps) to 512k sps, and finally, a {\em Symbol Sync} block implementing the Gardner algorithm for the time error detector and a polyphase filter bank as the interpolating resampler, downsampling the stream from 512K sps to (the original) 256K sps.
We remark that our flowcharts reflect as much as possible the standard modulation and demodulation chains implemented in commercial off-the-shelf devices. At the same time, our flowcharts do not expose (or hide) features that might help or prevent the \ac{RFF} process. For example, the {\em Costas Loop} block---a standard component in any receiver---allows carrier recovery and TX-RX clocks synchronization, but it prevents (non-standard) solutions like~\cite{bechir_2024_tmlcn}, which exploit \ac{CFO} to perform \ac{RFF}. Finally, we consider the \ac{SNR} as our main control variable, so to make our results independent of the power levels and the geometry of the deployment. 

{\bf Communication link.} We consider both a wired link (RF cable) and a wireless link with two Vert900 antennas (Fig.~\ref{fig:cable_scenario}). We also use a signal attenuator of 30dB between the transmitter and the receiver, so as to understand the impact of wireless channel propagation and multipath fading on the anonymization process. In fact, as introduced in Sect.~\ref{sec:background}, multipath fading represents yet another source of noise potentially affecting \ac{RFF} performance.

{\bf Dataset collection.} We collected two datasets for a total of approximately 167 GB of data, available at~\cite{dataset_anon}. \emph{The artifacts uploaded on the submission platform consist of all the post-processed data, processing scripts and intermediate data to generate the figures. The raw data (167GB) are too large to be shared anonymously. Thus, they will be made public at a later moment.} The dataset collected from the wired link ($\approx$ 134GB) includes 240 measurements, i.e., the result of 10 transmitters, 4 types of noise (no noise, Gaussian, Impulse, Laplacian, and Uniform), and 6 noise levels. The dataset collected from the wireless link ($\approx$ 33GB) includes 60 measures, i.e., the result of 10 transmitters, 1 types of noise (Gaussian), and 6 noise levels. Each measurement lasts for 300 seconds, thus providing 750M of \ac{IQ} samples (sample rate 250k sps).

\section{Device fingerprinting}
\label{sec:device_fingerprinting}
In this section, we present the \ac{RFF} techniques considered for analysis. We recall that mitigating multipath fading is critical for improving \ac{RFF} performance, and several techniques have been proposed, e.g., preprocessing of the \ac{IQ} samples, adaptation of (only) the first layers of the neural network, up to designing from scratch neural networks able to handle \ac{IQ} samples (see Sec.~\ref{sec:related_work}). Nevertheless, training and testing (directly) on \ac{IQ} samples require adapting the neural network (at least for the initial layers) and an ad-hoc preprocessing technique to mitigate the impact of the multipath fading. 
Recent state-of-the-art contributions, e.g., PAST-AI~\cite{oligeri2023tifs}, show the effectiveness of preprocessing \ac{IQ} samples into images and then using state-of-the-art \ac{DL} techniques to classify images. The root idea consists of generating images from \ac{IQ} samples while mitigating the impact of the noise by averaging the spatial position of the \ac{IQ} samples in the \ac{IQ} plane. As \acp{CNN} are a well-known class of \ac{DL} neural networks specifically designed to classify images, image-based RFF techniques use \acp{CNN} to discriminate images generated from samples of a specific transmitter, thus performing device fingerprinting.
We recall below the most important steps associated with the considered methodologies. Interested readers can refer to the relevant literature for details.

\begin{figure}[t]
    \centering
    \vspace{-18mm} 
    \subfloat[\centering \ac{IQ} samples]{{\includegraphics[width=4cm, angle=0, trim= 30mm 80mm 30mm 20mm,clip]{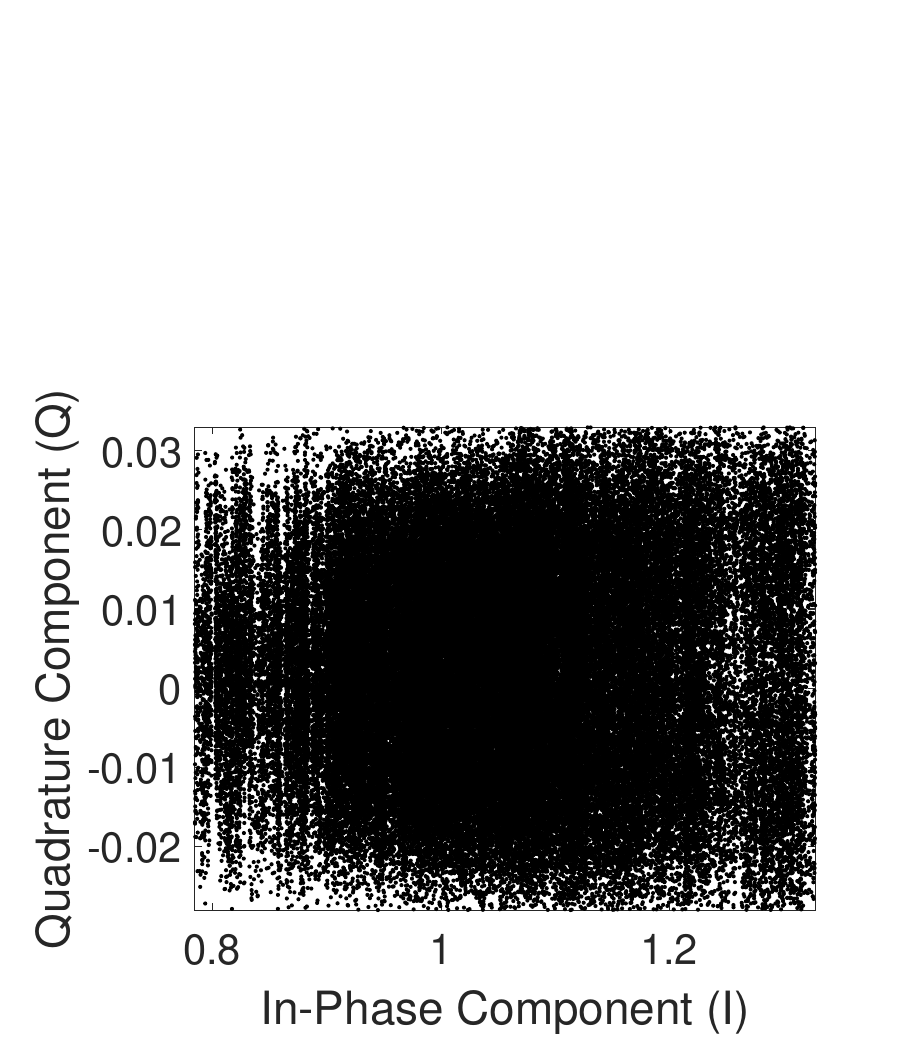} }}%
    \subfloat[\centering Bi-variate histogram]{{\includegraphics[width=4cm, angle=0, trim=30mm 75mm 30mm 100mm,clip]{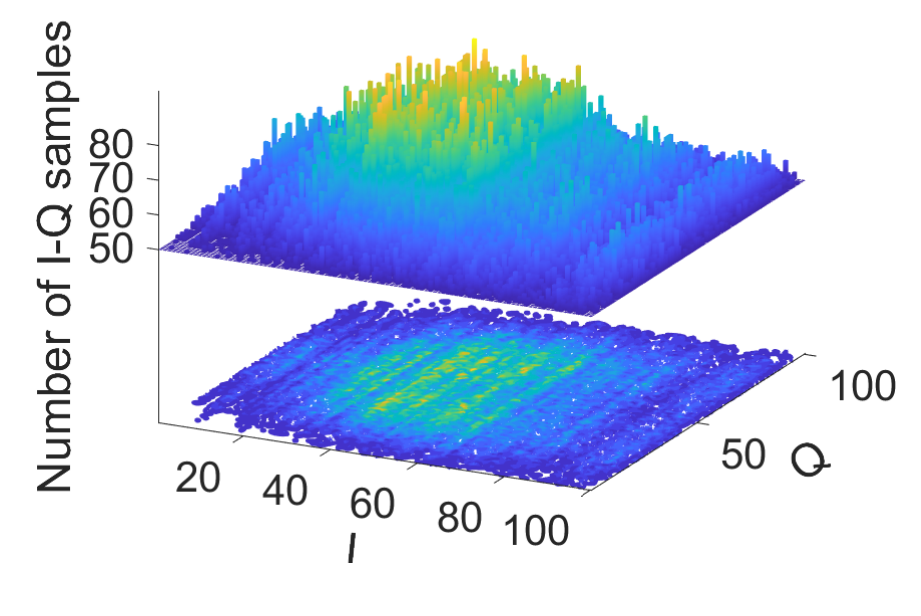} }}%
    \caption{We split the data collected (\ac{IQ} samples) into chunks of $10^5$ samples and organized them in one only cloud, i.e., the left cloud of the BPSK modulation is mirrored on the right side. We then compute a bi-variate histogram and consider the output as an image, ready to be processed by a \ac{DL} algorithm.}
    \label{fig:IQtoImage}%
\end{figure}

{\bf Data pre-processing.} Our input is a matrix of 2 columns (\ac{IQ} components) and 750M rows, one row for each \ac{IQ} sample. As depicted in Fig.~\ref{fig:solutionIQ}, the vast majority of the \ac{IQ} plane is not used, and therefore we split the \ac{IQ} plane into two parts (the positive and negative In-phase components) and then cut out the 0.005 and 0.0095 quantiles of each axis. Considering the reference value of $10^5$ \ac{IQ} samples per image (chunks extracted from the measurement), such a strategy involves removing a total of $2,000$ samples per chunk, allowing us to center the image on each cloud of \ac{IQ} samples while removing outliers (Fig.~\ref{fig:IQtoImage}(a)). The subsequent step consists of computing the bi-variate histogram on the data from Fig.~\ref{fig:IQtoImage}(a), as depicted in Fig.~\ref{fig:IQtoImage}(b). The number of bins of the bi-variate histogram is defined according to the size (width, height) of the images to be generated, and in turn, according to the input layer of the \ac{CNN}. As discussed in the following, we consider the \ac{CNN} {\em ResNet-18}, already implemented in Matlab2023b, thus setting the image size to 224-by-224 pixels. Figure~\ref{fig:IQtoImage}(b) wraps up on the performed computations: the bi-variate histogram counts the \ac{IQ} samples per bin (i.e., the density), and such value is considered as the value of the pixel in the generated image (lower part of Figure~\ref{fig:IQtoImage}(b)). The number of \ac{IQ} samples considered per image ($10^5$) should be carefully evaluated. Indeed, a high number of \ac{IQ} samples per image might generate too many tiles of the bi-variate histogram characterized by values bigger than 255, which is a non-consistent pixel value, thus leading to significant information loss. We carefully estimated the value of the number of \ac{IQ} samples per image, i.e., $10^5$, with empirical testing and in accordance with~\cite{oligeri2023tifs,alhazbi2023_acsac}.

{\bf Deep Learning.} We consider two \ac{DL} architectures, i.e., \ac{CNN} and \ac{AE}, as a function of the attack performed by the adversary (recall Sect.~\ref{sec:adversarial_model}). When considering the adversary $\mathcal{A}_1$, we assume the adversary targets a finite pool of devices and has already trained a model on the full set. Each device in the pool is willing to keep its anonymity at maximum (k-anonymity, as per the definition in~\cite{Samarati1998ProtectingPW}), i.e., accuracy equal to 0.1, with $k = 10$ devices in the pool.
When considering $\mathcal{A}_2$, we consider a targeted attack, i.e., the adversary trains his model on a dataset coming from a specific device, and subsequently, he wants to detect the presence of such a device in the radio spectrum. This problem can be addressed as a one-class classification problem, where the samples from the test set are detected as either \emph{legitimate}, if they are recognized to be coming from the target device, or \emph{anomaly}, if they are coming from any other device. 
Our reference \ac{CNN} {\em ResNet-18} is already implemented in Matlab R2023b and pre-trained on ImageNet~\cite{deng2009imagenet}, in accordance with the \emph{Transfer Learning} design principle. The main objective of this work is not to find the best configuration for the \ac{RFF} task, which has been addressed in previous work~\cite{alhazbi2023_acsac}. We choose {\em ResNet-18} in line with~\cite{oligeri2023tifs}, as it represents a good trade-off between accuracy and training time. Moreover, {\em ResNet-18} cannot be used as it is: its last layer requires adaptation to the number of classes (transmitters) in our problem, i.e., 10. Finally, we partially retrained the network by considering a stop criterion according to which the training is interrupted when the accuracy during the validation process achieves a variance across the last 5 iterations less than 0.5.

Our reference {\em autoencoder}, in line with~\cite{oligeri2023tifs}, is characterized by 64 neurons in the hidden layer, the logistic sigmoid function as encoder transfer function, a decoder transfer function implemented with a linear transfer function, a sparsity regularization coefficient of 0.5, and finally, a $L_2$ weight regularizer coefficient equal to 0.01.
The autoencoder is constituted by an encoder and a decoder, and it outputs a reconstruction of the input. During the training phase, the encoder learns a set of features (latent representation), which is then used by the decoder to reconstruct the input. This makes the autoencoder applicable in scenarios where a prediction of inputs not previously seen is requested. Given the structure of the autoencoders, a fundamental metric is represented by the reconstruction error since it represents how the reconstructed output is ``far away'' from the input when reconstructed with the input's features. In the remainder of this work, we will use the \ac{MSE} to assess if the image (and the associated \ac{IQ} samples) belongs to the purported transmitter. An image with $MSE < \tau$ is predicted to be generated from the target transmitter, while an image with $MSE > \tau$ is predicted to be generated from another transmitter. For computing $\tau$, we adopt the standard formula in Eq.~\ref{eq:Threshold_Formula}, where $E(\circ)$ and $\sigma(\circ)$ represent the mean and the standard deviation, respectively, of the \ac{MSE} obtained during the validation process.
\begin{equation}
    \label{eq:Threshold_Formula}
    \tau = E(MSE_{val}) + 3.5 \cdot \sigma(MSE_{val}),
\end{equation}

\section{Results}
\label{sec:results}
In this section, we present the results of our analysis related to two models of adversaries, i.e., $\mathcal{A}_1$ and $\mathcal{A}_2$ (recall Sect.~\ref{sec:adversarial_model}), while considering an increasing a priori knowledge of the transmitters. Indeed, we can make two different assumptions on the background knowledge of the adversary, independently of the model. We can assume that the adversary was able to train a model either on noise-free samples or on a combination of noise-free and noisy samples (all samples). 

\subsection{Wired Link - Adversary $\mathcal{A}_1$}
\label{sec:adversary_a1}
\begin{figure}[t]
    \centering
    \includegraphics[width=\columnwidth, angle = 0,trim = 30mm 80mm 30mm 90mm]{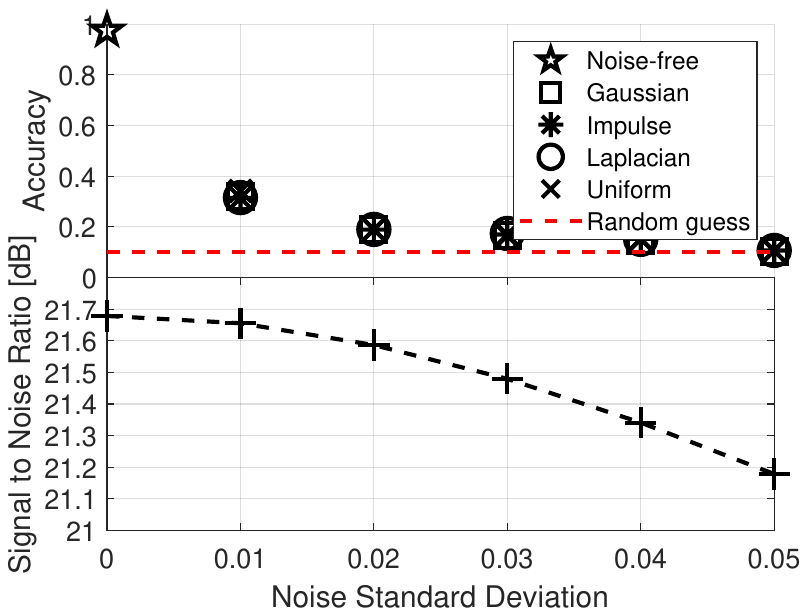}
    \caption{Adversary $\mathcal{A}_1$ training on noise-free samples and testing on measurements with noise greater than zero. The adversary trains a model with 10 classes (one for each transmitter) on measurements with no noise and then is challenged to identify the transmitter.} 
    \label{fig:cleanVSnoisy}
\end{figure}

\begin{figure}[t]
    \centering
    \subfloat[\centering Noise-free]{{\includegraphics[width=4cm, angle=0, trim= 35mm 85mm 35mm 90mm]{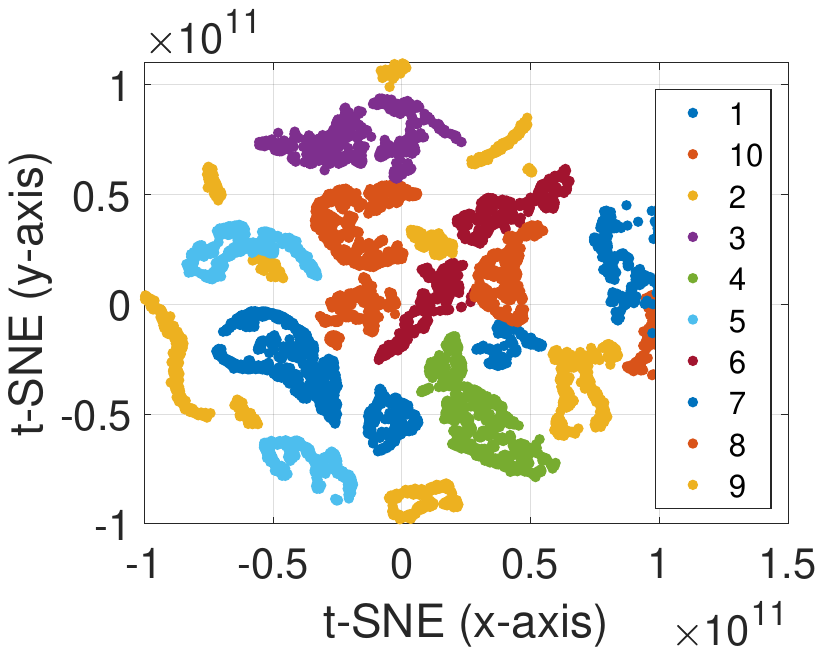}}}
    \subfloat[\centering Noise level 0.02]{{\includegraphics[width=4cm, angle=0, trim= 35mm 85mm 35mm 90mm]{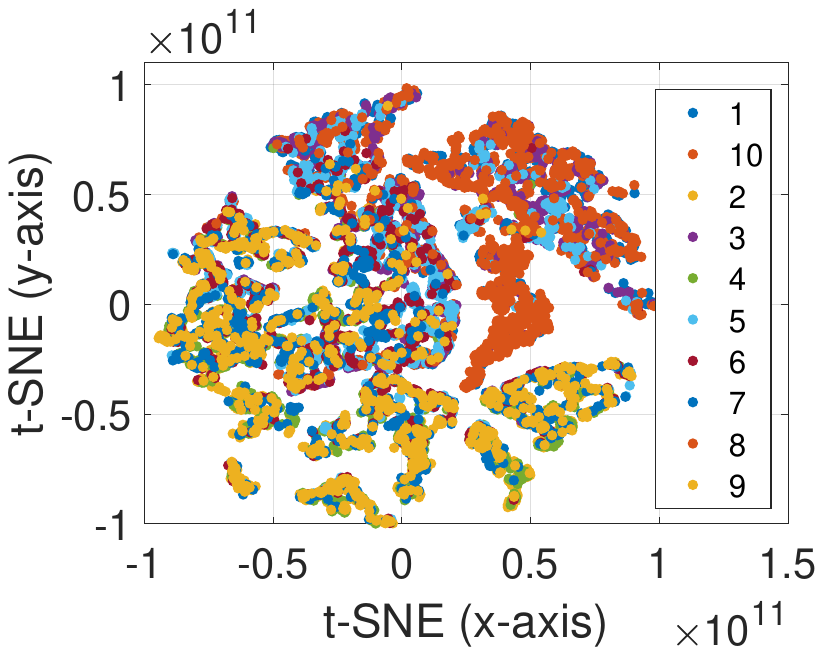}}}
    \caption{t-SNE analysis for noise-free and $\sigma=0.02$. Classification is possible with no noise, but adding noise makes the transmitters virtually indistinguishable when $\sigma \ge 0.02$.}
    \label{fig:sne}
\end{figure}

{\bf Training on noise-free samples.} We first consider the measurements acquired using the wired communication link and focus on training the {\em ResNet-18} \ac{CNN} with noise-free data and testing on any noise level. As previously introduced in Sect.~\ref{sec:measurement_setup}, for each device, we collected six different noise levels, i.e., standard deviation $\sigma = \{0, 0.01, 0.02, 0.03, 0.04, 0.05\}$, considering five different noise models, i.e., noise-free, Gaussian, Impulse, Laplacian, and Uniform. We observe that $\sigma$ is dimensionless, since both I and Q components are normalized to be placed at the ideal positions in the IQ plane as per Fig.~\ref{fig:solutionIQ}. Therefore, we considered 10 classes (one for each transmitter) and trained a model using the 10 available measurements with noise levels equal to zero. Subsequently, we tested our model considering all combinations of transmitters and noise levels (greater than zero). Figure~\ref{fig:cleanVSnoisy} shows the results of our analysis. In particular, the upper part of Fig.~\ref{fig:cleanVSnoisy} depicts the accuracy of the classifier as a function of the noise level ($\sigma$) while considering all the noise types provided by the GNURadio block {\em Noise Source}. We observe that the classification accuracy is not affected by the noise type, since they all exhibit the same behavior. In contrast, the accuracy is strongly affected by the noise standard deviation $\sigma$. Starting from accuracy $0.99$ with $\sigma=0$ (baseline), the value $\sigma \approx 0.02$ is enough to achieve the random guess (accuracy close to 0.1). In the bottom part of Fig.~\ref{fig:cleanVSnoisy}, we considered the \ac{SNR} as a function of the noise level introduced by the transmitter. We highlight that a variation in the \ac{SNR} of about 0.1dB is enough to achieve anonymity, thus proving that the quality of the link is negligibly affected by the injection of noise. For the sake of completeness, we recall Fig.~\ref{fig:solutionIQ}, which has been generated by considering $10^5$ samples from the noise-free configuration and $10^5$ samples from the configuration combining Gaussian noise and $\sigma = 0.05$.
Finally, we consider the t-Distributed Stochastic Neighbor Embedding (t-SNE) analysis associated with two cases, i.e., noise-free and $\sigma = 0.02$, as shown in Fig.~\ref{fig:sne}. The t-SNE analysis performs a dimensionality reduction, i.e., from 10 activations extracted from the last layer of the {\em ResNet-18} network to a two-dimensional space that can be represented in a Cartesian plane. As shown in Fig.~\ref{fig:sne}(a), noise-free values are very well disjoint, and clustering makes it possible to assign each point (image) to the related class. This is not possible in the scenario of Fig.~\ref{fig:sne}(b), where $\sigma = 0.02$ leads to the overlap of the images of the different classes.

\begin{figure}[t]
    \centering
    \subfloat[\centering Confusion Matrix]{{\includegraphics[width=4cm, angle=0, trim= 40mm 85mm 40mm 90mm]{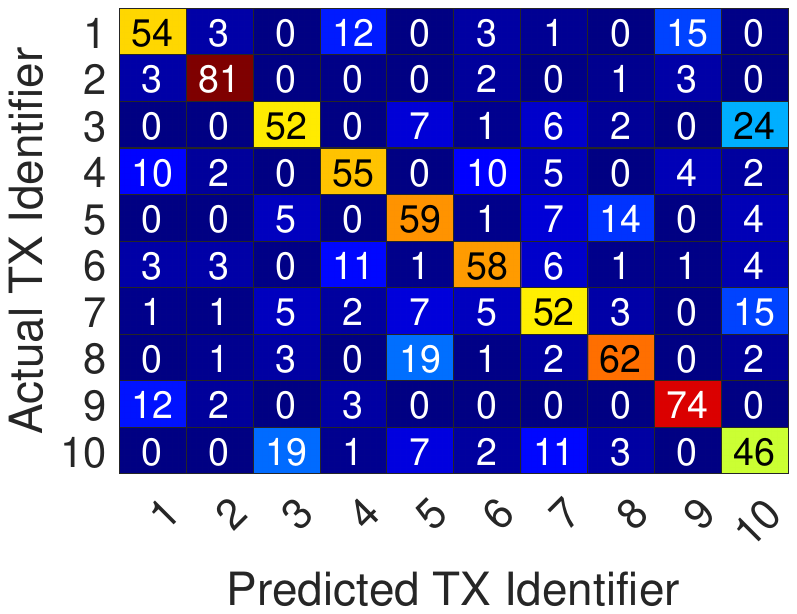} }}
    \subfloat[\centering FPR-FNR]{{\includegraphics[width=4cm, angle=0, trim= 40mm 85mm 40mm 90mm]{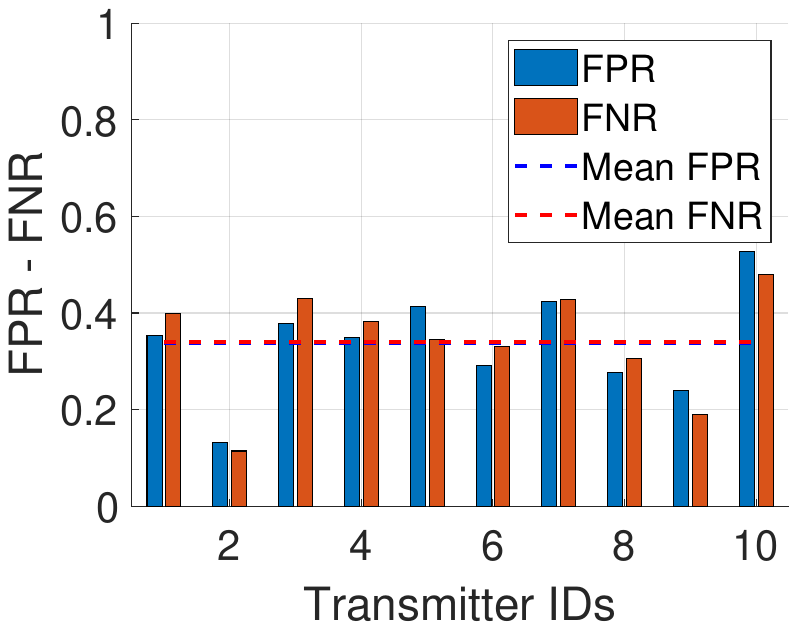} }}
    \caption{Adversary $\mathcal{A}_1$ training on all noise levels and testing with random samples taken from random measurements with noise strictly greater than zero. The adversary trains a model with 10 classes (one for each transmitter) on measurements with all the noise levels spanning between 0 and 0.05, and then it is challenged to identify the transmitter.}
    \label{fig:cnn_all}
\end{figure}

{\bf Training on all samples.} We also consider the scenario where the adversary trains the {\em ResNet-18} \ac{CNN} on all the (10) transmitters in the pool and noise levels while considering only the Gaussian noise model---recall from Fig.~\ref{fig:cleanVSnoisy} that different types of noise have the same impact on the fingerprint. To achieve this, we consider the whole dataset and split it into three disjoint chunks, i.e., 60\%, 20\%, and 20\%, for training, validation, and testing, respectively. This adversary configuration is the most powerful considered in this work. Indeed, as the most favorable condition for the adversary, we assume that the adversary is able to collect (and label) samples from all the transmitters with all the noise levels. Figure~\ref{fig:cnn_all}(a) shows the confusion matrix associated with the actual and predicted transmitters. We observe that, with such knowledge, $\mathcal{A}_1$ significantly increases its chances of detecting the presence of a transmitter of the pool in the radio spectrum (the average accuracy is about 0.66). Nevertheless, our solution still hides the identity of the transmitters, as depicted in Figure~\ref{fig:cnn_all}(b). Indeed, the average \ac{FPR} and \ac{FNR}, computed on all the transmitters, are similar and equal to 0.35 (dashed red and blue line in Fig.~\ref{fig:cnn_all}(b)). A significant number of transmitters (6 out of 10) are above the average, meaning that when the classifier predicts a specific transmitter, the probability of an error is higher than 0.35. We stress that this value has been obtained by giving to the adversary all the possible knowledge available in the system. Moreover, recalling that the transmitter and the receiver are connected by an RF cable, we highlight that this setup avoids the multipath fading, which in turn, helps as well to hide the fingerprint, thus representing an upper bound on the performance of the adversary.

\subsection{Wired Link - Adversary $\mathcal{A}_2$}
\begin{figure}[t]
    \centering
    \includegraphics[width=\columnwidth, angle = 0,trim = 40mm 80mm 40mm 85mm]{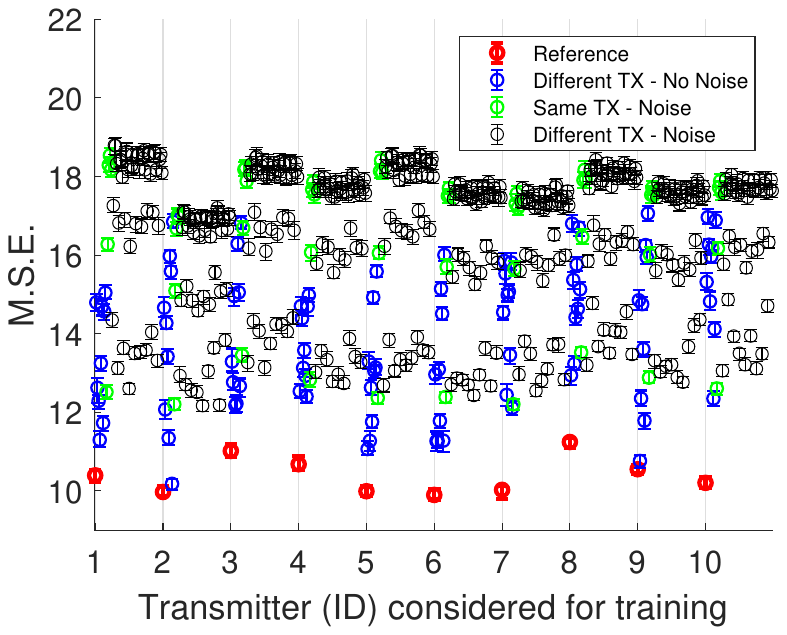}
    \caption{The adversary $\mathcal{A}_2$ trains 10 different models, one for each transmitter, considering only the noise-free samples. 
    Red bars represent the test results performed on the same measurements considered for training (split ratio 80/20). Subsequently, the adversary challenges the trained model with additional datasets: (i) noise-free samples from different transmitters (blue bars), (ii) noisy samples from the same transmitter considered for the training (green bars), and finally, (iii) noisy samples from different transmitters (black bars).} 
    \label{fig:autoencoders}
\end{figure}

{\bf Training on noise-free samples.} We still consider the measurements acquired using the wired link and consider the case of training on noise-free samples for a specific device (only), in line with the adversary model $\mathcal{A}_2$. Then, we tested on all the available transmitter configurations, i.e., noise-free and with Gaussian noise with standard deviation $\sigma = \{0, 0.01, 0.02, 0.03, 0.04, 0.05\}$. We highlight that we did consider only one noise model (Gaussian) in this case since the noise pattern does not affect the accuracy of the classification process (recall Sec.~\ref{sec:adversary_a1} and Fig.~\ref{fig:cleanVSnoisy}). Figure~\ref{fig:autoencoders} shows the results of our analysis in terms of \ac{MSE} as a function of the transmitter identifier (ID) used for training. Each bar represents the quantiles 5, 50, and 95 associated with the \ac{MSE} computed on the images generated from the \ac{IQ} samples. We highlight the comparisons we have made considering four colors: (i) the comparison of the reference set, i.e., two disjoint datasets (60\% and 40\%) taken from each transmitter ID (red color), one used for training and another one used for testing, respectively; (ii) the comparison of the reference transmitter ID and any other transmitter ID in the pool while considering noise-free samples (blue color), (iii) the comparison of each transmitter ID with the same ID but considering the measurements with the Gaussian noise (green color), and finally, (iv) the comparison of the reference transmitter ID with any other transmitters but considering the measurements with the noise (black color). The red bars in Fig.~\ref{fig:autoencoders} serve as the reference values. We selected 10 noise-free measurements from each transmitter and then split the generated images into two datasets (60\% and 40\%) for training and testing, respectively. This setup provides the \ac{MSE} reference values to be compared with the images coming from other configurations.
When comparing the red and blue bars (different transmitter IDs and noise-free measurements), there is no (major) overlap that proves that transmitters can be distinguished in a noise-free environment. The minor overlap between transmitters 2 and 9 will be investigated later on. An orthogonal analysis involves the performance estimation of our solution by comparing the noise-free measurements (red bars) with the ones affected by \sol\ (green bars), where we injected Gaussian noise with different standard deviations. The injected noise significantly affects the \ac{MSE}, and in particular, it makes blue bars overlap with the green ones, i.e., the reference transmitter becomes indistinguishable from the others. Finally, the black error bars represent the measurements with noise from all the transmitters. Note that the noise makes all the measurement configurations overlap, thus achieving transmitter indistinguishability. 

\begin{figure}[t]
    \centering
    \includegraphics[width=0.75\columnwidth, angle = -90,trim = 0mm 0mm 0mm 0mm]{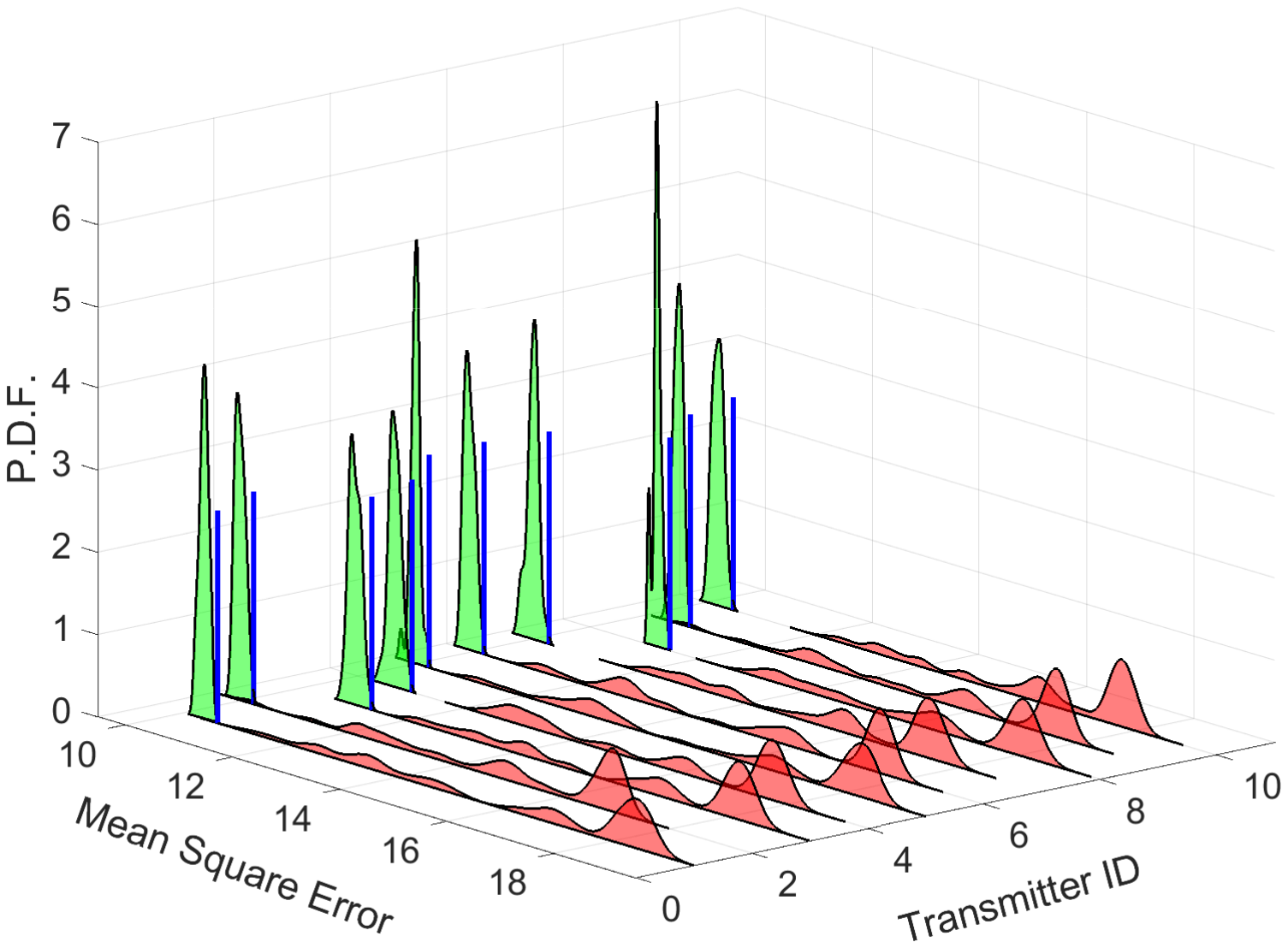}
    \caption{\ac{MSE} values when training on noise-free samples (green curves) and testing on the rest of the dataset (red curves). The blue lines are the thresholds computed via Eq.~\ref{eq:Threshold_Formula}. The overlap between the two sets of distributions (red and blue) is negligible for all the transmitters: the autoencoder is successful in distinguishing the transmitters.} 
    \label{fig:autoenc_thr}
\end{figure}

To quantify the accuracy that can be achieved considering the attack deployed by adversary $\mathcal{A}_2$, for each transmitter, we define a decision threshold according to Eq.~\ref{eq:Threshold_Formula}. Figure~\ref{fig:autoenc_thr} shows the results of our analysis while comparing noise-free signals (green shaded areas) with the remainder of our dataset (red shaded areas). We highlight in blue the thresholds computed for each transmitter. Our analysis shows that the fingerprint (completely) changes when noise is added to the transmitted signal. Indeed, the \ac{FPR}, i.e., samples from noise-free measurements leading to MSE values larger than the threshold values, is zero independently of the considered transmitter. The false negative ratio, i.e., samples taken from any other measures, is always zero except for two cases, which are about 0.01. This is the case of the transmitters 2 and 9, which have a similar fingerprint (recall Fig.~\ref{fig:autoencoders}). Finally, recalling Fig.~\ref{fig:autoencoders}, we observe that different configurations (noise-free and noisy samples) eventually collapse in the same range (red shaded areas in Fig.~\ref{fig:autoenc_thr}), thus making the detection of each of them challenging. For example, an adversary might be able to collect noisy samples with a specific noise level, then train a model, and finally accept the challenge of detecting such a pattern in the radio spectrum. This is unlikely, since all comparisons fall in the region with \ac{MSE} values between 11 and 19, and therefore, the adversary would experience a high number of misclassifications independently of the chosen configuration.

\begin{figure}[t]
    \centering
    \includegraphics[width=0.75\columnwidth, angle = -90,trim = 0mm 0mm 0mm 0mm]{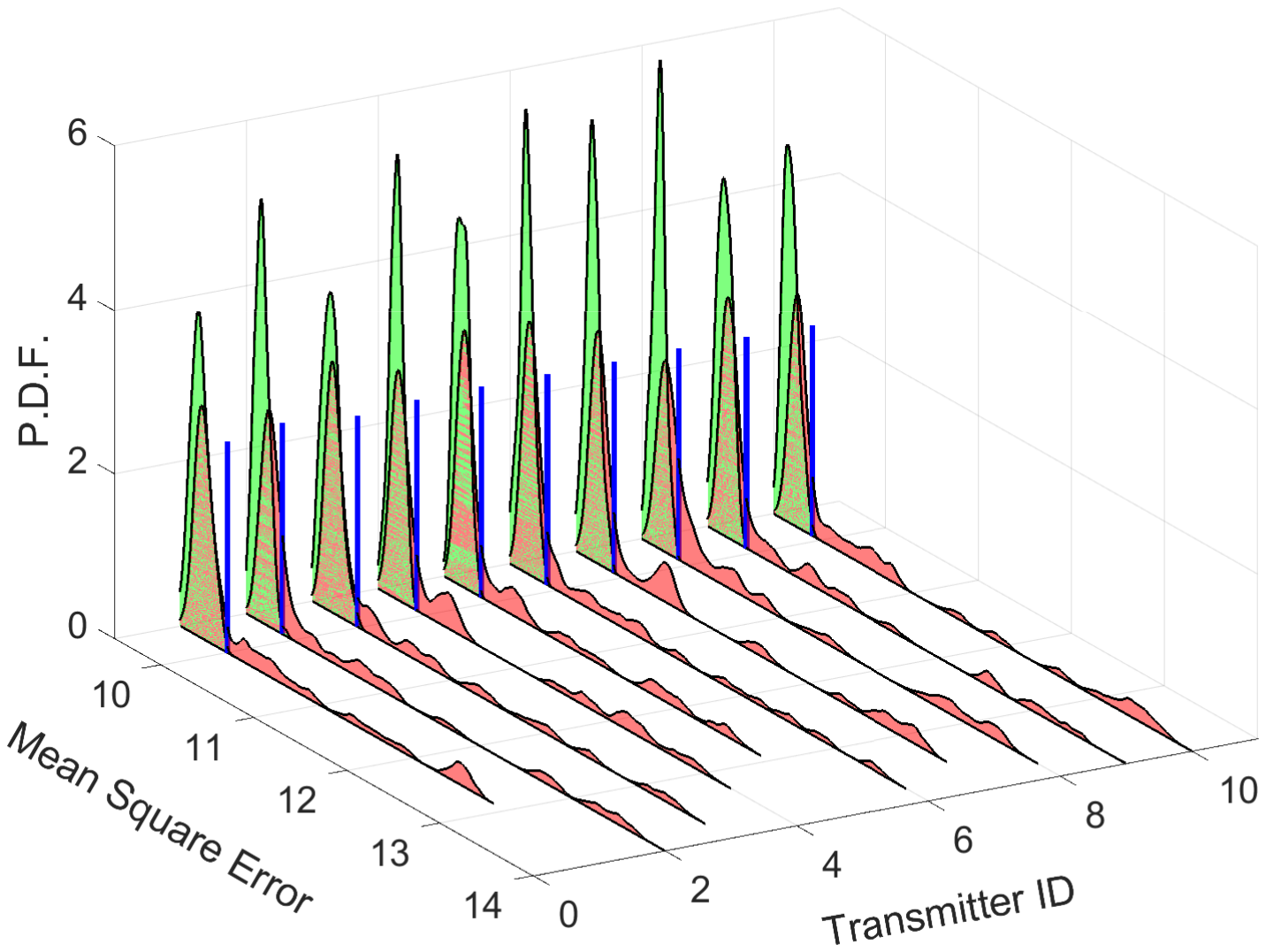}
    \caption{\ac{MSE} values when training on any samples (green curves) and testing on the rest of the dataset (red curves). The blue lines are the thresholds computed via Eq.~\ref{eq:Threshold_Formula}. The overlap between the two sets of distributions (red and blue) is not negligible for all the transmitters (0.66 on average), and thus, the autoencoder fails to distinguish the transmitters.} 
    \label{fig:autoenc_thr_noisy}
\end{figure}

{\bf Training on all samples.} In the following, we consider the case of adversary $\mathcal{A}_2$ owning a model trained on both noise-free and noisy samples. Indeed, the adversary has been exposed to the full dataset (noise-free and noisy samples) associated with each device considered in our test. As before, we considered the Gaussian noise with standard deviation \mbox{ $\sigma = \{0, 0.01, 0.02, 0.03, 0.04, 0.05\}$}. Figure~\ref{fig:autoenc_thr_noisy} shows the results of our analysis, where we perform a targeted attack (adversary $\mathcal{A}_2$) on each device of the pool (transmitter IDs). We report, for each of them, the \ac{MSE} associated with the test set from the selected transmitter ID (green shaded area) and the \ac{MSE} associated with the other measurements (other transmitters with any noise level). The probability distribution functions overlap for all the transmitter IDs. When considering the threshold defined by Eq.~\ref{eq:Threshold_Formula} (blue lines in Fig.~\ref{fig:autoenc_thr_noisy}), we obtain an average false negative ratio of about 0.66, confirming that the autoencoder fails to detect the presence of a specific transmitter in the radio spectrum when the training is performed on all the available samples (noise-free and noisy). Such a finding is confirmed by the empirical intuition that a model trained on all the samples cannot differentiate a specific transmitter from the pool since the (injected) noise makes the fingerprint of all the transmitters very similar to each other.

\subsection{Wireless Link}
\label{sec:wireless}
\begin{figure}[t]
    \centering
    \includegraphics[width=\columnwidth, angle = 0,trim = 30mm 90mm 30mm 90mm]{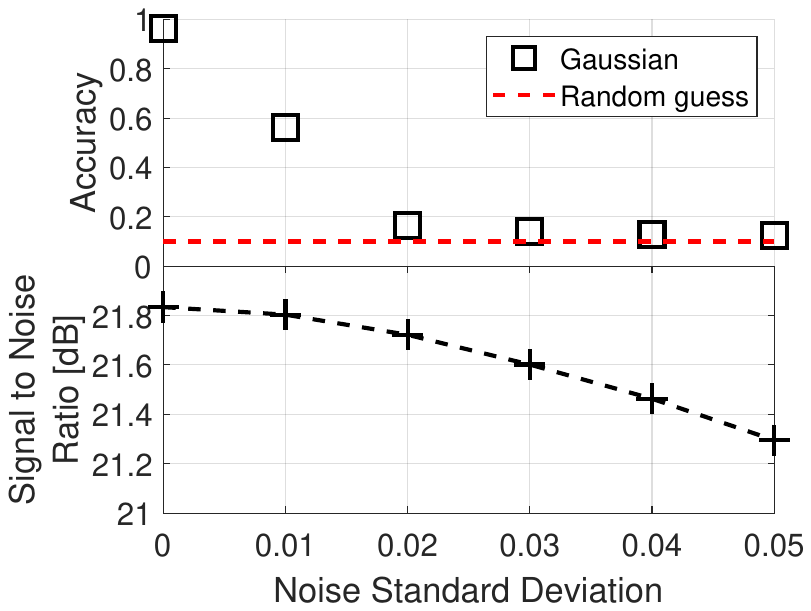}
    \caption{Adversary $\mathcal{A}_1$ training on noise-free samples from a wireless link. The adversary trains a model with 10 classes (one for each transmitter) on measurements with noise-free samples and then has to identify the transmitter(s) by considering measurements with noise strictly greater than zero.} 
    \label{fig:radio_cnn}
\end{figure}

Our wireless scenario involves the same configuration as the wired one (recall Fig.~\ref{fig:cable_scenario}). In line with the literature~\cite{alhazbi2023_acsac}, we placed the transmitters and the receiver a few meters away in an office during working time, with people moving around in close proximity generating multipath. We consider the same data collection procedure as that described for Fig.~\ref{fig:cleanVSnoisy}, i.e., Gaussian noise type, \ac{CNN} {\em ResNet-18}, and $10^5$ samples per image. \\
{\bf Adversary $\mathcal{A}_1$.} Figure~\ref{fig:radio_cnn} (top) shows the accuracy of {\em ResNet-18} as a function of the noise standard deviation $\sigma$. When $\sigma = 0$ (noise-free), the accuracy is approximately 0.96, while dropping to 0.16 when $\sigma = 0.02$. It is worth noting that an accuracy value of 0.1 is equivalent to a random guess. The \ac{SNR} is consistent with the one experienced when using the cable: we kept the attenuation to 30dB, and the distance between the transmitter and the receiver was only a few meters. Moreover, we observe the existence of a transient region in the range $]0, 0.02]$ where the accuracy is affected by the combination of different effects, i.e., the fingerprint of the device and the fingerprint of the noise introduced by the channel~\cite{sadighian2024ccnc,oligeri2024sac}. The measurements involving the RF cable are not affected by external factors (multipath), and therefore, the fingerprint of the device is the principal component of the received signal. In contrast, when transmitting over the wireless channel, the channel impulse response varies over time depending on the communication link. Therefore, the \ac{DL} model is likely to experience either consistent or different channel states (channel fingerprint). The final result, as observed by other contributions~\cite{sadighian2024ccnc,oligeri2024sac,alhazbi2023_acsac}, is that \ac{RFF} is affected by both the device and the channel fingerprint. The accuracy for $\sigma=0.01$ in Fig.~\ref{fig:radio_cnn} (top), i.e., 0.56, is higher than the one in Fig.~\ref{fig:cleanVSnoisy} (top), i.e., 0.25, and it can be explained considering the higher impact of the wireless channel. Indeed, the measurements taken considering different transmitters have the same cable with the same channel fingerprint, while this is not happening in the wireless scenario, where different transmitters might be affected by a (slightly) different radio channel, which helps the classification. Finally, we observe that a minimum noise standard deviation of 0.02 fully anonymizes the transmitter in any scenario. Moreover, the \ac{SNR} drop (0.1dB) necessary to remove the fingerprint is consistent with the one experienced using the cable.

\begin{figure}[t]
    \centering
    \includegraphics[width=\columnwidth, angle = 0,trim = 30mm 80mm 30mm 90mm]{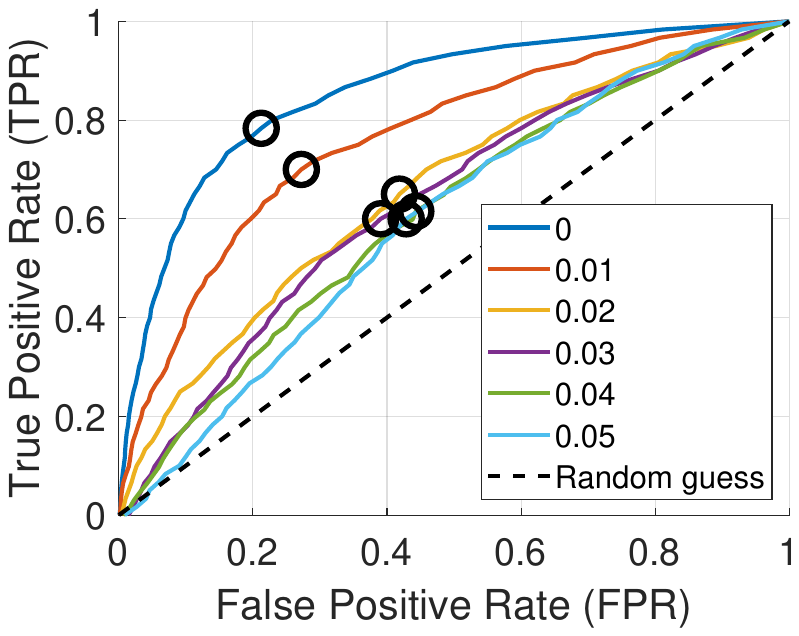}
    \caption{Adversary $\mathcal{A}_2$. We trained an autoencoder on measurements taken from a target device and tested against the others in the pool while keeping the same noise level $\sigma \in \{0, 0.01, 0.02, 0.03, 0.04, 0.05\}$ in an actual wireless link.} 
    \label{fig:autoenc_roc_wireless}
\end{figure}

{\bf Adversary $\mathcal{A}_2$.} We now consider adversary $\mathcal{A}_2$, i.e., a targeted attack against a specific transmitter in our pool of 10. We trained an autoencoder on wireless measurements taken from a specific device while setting the noise level $\sigma$ and tested it on the remainder of the measurements collected from the other transmitters in the pool, given the same $\sigma$. In fact, we assume $\mathcal{A}_2$ is capable of training from any type of measurement (noise-free and noisy) and then test it in the wild by collecting samples from the radio spectrum. Figure~\ref{fig:autoenc_roc_wireless} shows the results of our analysis, where we consider the average \ac{ROC} curves calculated considering the mean value over the 10 transmitters when varying $\sigma \in \{0, 0.01, 0.02, 0.03, 0.04, 0.05\}$. Our results confirm previous findings. First, we observe that the performance of the autoencoder in a noise-free environment (solid blue line in Fig.~\ref{fig:autoenc_roc_wireless})) is characterized by an FPR and TPR of 0.21 and 0.78, respectively, when considering the point in the \ac{ROC} curve that offers the best trade-off between sensitivity and specificity, under the assumption that both types of classification errors (TPR and FPR) cost the same---the overall accuracy is equal to 0.81. Moreover, a Gaussian noise with standard deviation $\sigma = 0.01$ is not enough to de-anonymize the transmitter. Indeed, the solid red line in Fig.~\ref{fig:autoenc_roc_wireless}, representing the \ac{ROC} curve for $\sigma = 0.01$, is characterized by FPR and TPR equal to about 0.27 and 0.7, respectively---the overall accuracy is equal to 0.73. Similarly to the previous case ($\mathcal{A}_1$), $\sigma = 0.02$  allows us to achieve a much better anonymization level, i.e., TPR and FPR values equal to 0.27 and 0.7, respectively, with an overall accuracy equal to 0.64.

\section{Comparison}
\label{sec:comparison}
The RFF methodology used in the previous section to validate \sol\ involves the pre-processing of \ac{IQ} samples into images, as described in Sect.~\ref{sec:device_fingerprinting}. As shown by the authors in~\cite{alhazbi2023_acsac}, this strategy is effective in mitigating the impact of noise on the \ac{RFF} process. To provide further insights and show that \sol\ is effective also considering other RFF approaches, in this section, we test our anonymization solution against a different approach considered in the literature, i.e., the use of (raw) \ac{IQ} samples to train and test a \ac{CNN}, as considered by~\cite{liu2023_wcl},~\cite{sun2022_rs}, and~\cite{lu2024_arxiv}. 
We highlight that for image-based pre-processing, we resort to the implementation of \ac{CNN} in MatLab that cannot be used out of the box, but it requires some modification to the input and output layers. It is worth mentioning the choice of using an already designed (and trained) network, despite designing and training it from scratch. Indeed, considering a self-designed network makes the comparison with the state-of-the-art even more challenging, since all the contributions adopting such a strategy are not consistent in the design of the neural network. Therefore, to provide a fair comparison, we consider {\em ResNet-18} and re-adapt it to accept as input chunks of $10^4$ \ac{IQ} samples. The size of the chunk has been evaluated as a trade-off between computation overhead and performance. 

\begin{figure}[t]
    \centering
    \includegraphics[width=\columnwidth, angle = 0,trim = 30mm 80mm 30mm 90mm]{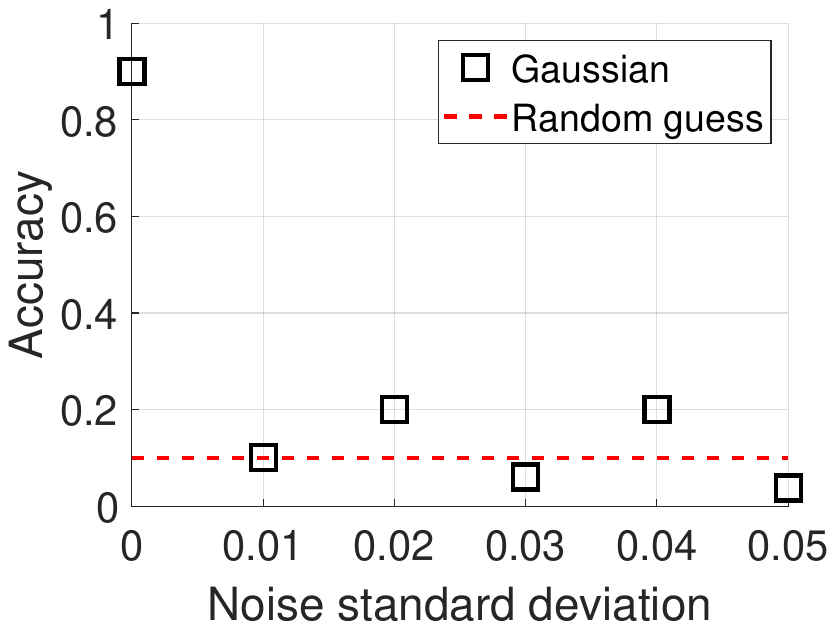}
    \caption{RFF Performance when considering raw \ac{IQ} samples and different levels ($\sigma$) of Gaussian noise. Standard techniques using raw \ac{IQ} samples for training and testing are more sensitive to noise injection, thus allowing perfect anonymization even for $\sigma = 0.01$.}  
    \label{fig:rawiq_cnn_wireless}
\end{figure}

We consider the dataset collected in the wireless scenario, training on noise-free samples and then testing on increasing values of $\sigma$, with Gaussian noise. Figure~\ref{fig:rawiq_cnn_wireless} shows our results. We observe that even with the smallest noise level ($\sigma=0.01$), the accuracy of RFF via raw \ac{IQ} samples drops to the random guess (0.1). Figure~\ref{fig:rawiq_cnn_wireless} should be compared with Fig.~\ref{fig:radio_cnn}, i.e., same scenario but different methodology (with and without pre-processing into images). We observe that both techniques have similar performance with noise-free samples ($\sigma = 0$), but image pre-processing is confirmed to be more reliable with noise: when considering $\sigma = 0.01$, the accuracy in Fig.~\ref{fig:radio_cnn} (image-based \ac{RFF}) is higher than in Fig.~\ref{fig:rawiq_cnn_wireless} (raw \ac{IQ}), although not acceptable for \ac{RFF} due to too high mispredictions ($> 0.4$). For higher values of $\sigma$, performance across \ac{RFF} techniques are the same and confirm the effectiveness of our solution.

\section{Selective Fingerprint Disclosure}
\label{sec:hideprint}
Our results (Sect.~\ref{sec:results}) show that the adversary (both $\mathcal{A}_1$ and $\mathcal{A}_2$) does not gain any advantage in adding additional noisy samples to the training process. In such scenarios, as shown in Fig.~\ref{fig:cleanVSnoisy}, the adversary experiences the disappearance of the fingerprint when the transmitters add an increasing level of noise. Nevertheless, in this section, we investigate the possibility of performing \emph{Selective Radio Fingerprint Disclosure}, i.e., allowing a legitimate receiver to successfully perform \ac{RFF} even when transmitters are adding noise to the signal, while avoiding the same for illegitimate parties (adversary).

\begin{figure}[t]
    \centering
    \vspace{-17mm} 
    \subfloat[\centering Confusion Matrix]{{\includegraphics[width=4cm, angle=0, trim= 40mm 85mm 40mm 40mm]{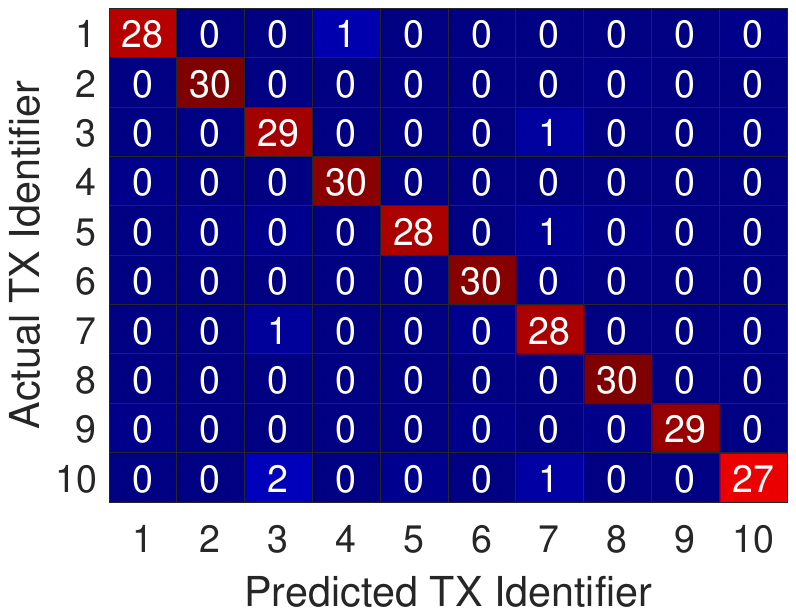} }}
    \subfloat[\centering FPR-FNR]{{\includegraphics[width=4cm, angle=0, trim= 40mm 85mm 40mm 40mm]{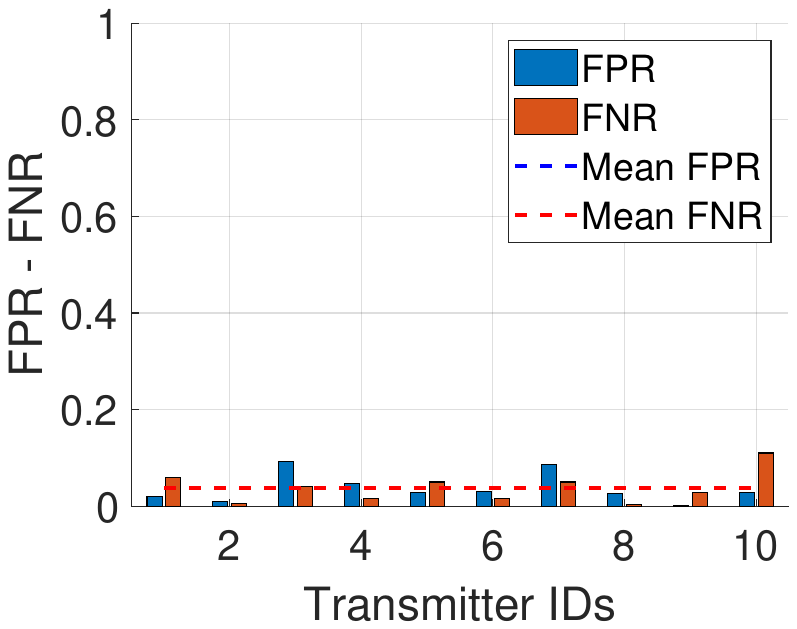} }}
    \caption{Selective Radio Fingerprint Disclosure. Only the receiver aware of the seeds $s_k$ can successfully perform \ac{RFF}.}
    \label{fig:hideprint}
\end{figure}

We consider a pool of $N$ devices, each of them being able to select a noise level in a pool of $M$ levels. Our following analysis shows that, under standard conditions, an adversary cannot infer the identity of the transmitter from all the available possibilities $[N \times M]$. In the following, we show that we can achieve such a property by sharing with a legitimate receiver only the noise level currently adopted while keeping it secret from the adversary. In this way, a legitimate receiver can authenticate the transmitter, but the adversary cannot.
We consider that the $N$ transmitters are loosely time-synchronized with the receiver, and that the time $t$ is divided into slots of duration $T$, i.e., $t:[t_0, \ldots, t_\infty]$. During each time slot, a transmitter can perform at most one communication~\cite{dipietro_2017_cn}, choosing the standard deviation $\sigma$ of the noise level ($n_i$) to be added to the signal via Eq.~\ref{eq:hideprint}.
\begin{equation} 
    n_i = H(s_k\ |\ t_i) \mod M,
    \label{eq:hideprint}
\end{equation}
being $H(\circ)$ a cryptographically secure hash function, $s_k$ the seed (secret to the adversary) preloaded on each transmitter with ID $k$, and $n_i$ the noise level to be used at time slot $t_i$. Each transmitter, at each slot $t_i$, computes the noise level to be applied to the radio signal according to a pseudo-random sequence that is unknown to the adversary--- the seed $s_k$ is secret to the adversary. The legitimate receiver, aware of the seed $s_k$, can compute the noise level adopted by the transmitter in the current time slot. Under this assumption, the legitimate receiver has a knowledge advantage with respect to the adversary, and can use the appropriate \ac{RFF} model with respect to the one of the adversary---which is the best that can be adopted under the considered assumptions, i.e., $\mathcal{A}_1$ considering all samples. Specifically, in the slot $t_i$, the receiver can use a model trained on only 10 classes, i.e., one per transmitter, each for a specific noise pattern. Any other (unauthorized) receivers, unaware of the secret $s_k$, do not know the noise level used by each transmitter, and has to use a model trained only on noise-free samples, which is more prone to misclassification than the one used by the legitimate receiver (see Fig.~\ref{fig:radio_cnn}). 
We acknowledge that such a strategy requires the legitimate receiver to pre-train ${N \choose M}$ models (for each transmitter, all possible noise levels). However, the training can be done offline. As an example, considering the configuration of this work, the receiver should first pre-train ${10 \choose 6} = 210$ models in the offline phase. At runtime, the receiver can pick the correct model according to the time slot $t_i$, using Eq.~\ref{eq:hideprint}, and finally test the samples collected from the radio spectrum.
To prove the effectiveness of such a solution against the adversary, we perform Monte-Carlo simulations (100 iterations) by considering a receiver aware of the noise levels considered by the pool of the transmitters. Figure~\ref{fig:hideprint} shows the results of our analysis. Our test set consists of (on average) 30 images per transmitter, each image being the result of $10^5$ \ac{IQ} samples. The average accuracy from Fig.~\ref{fig:hideprint}(a) exceeds 0.96, while we reported the \ac{FPR} and \ac{FNR} in Fig.~\ref{fig:hideprint}(b). The performance in Fig.~\ref{fig:hideprint} should be compared with those in Fig.~\ref{fig:cnn_all}, which is based on the logic used by the adversary. The knowledge of the noise currently adopted (available on a legitimate receiver) increases the accuracy from 0.66 to 0.96, allowing the receiver to reliably authenticate the transmitter.

\begin{figure}[t]
    \centering
    \includegraphics[width=\columnwidth, angle = 0,trim = 30mm 80mm 30mm 90mm]{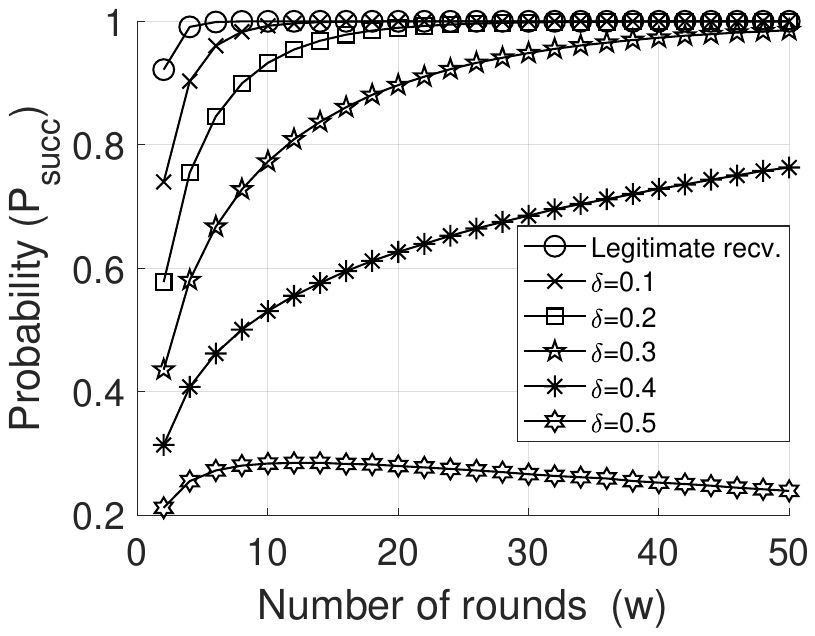}
    \caption{Probability of correct transmitter authentication ($P_{succ}$) with various rounds (time-slots). A legitimate receiver has an advantage (the knowledge of the seed $s_k$) that can be exploited to maximize its probability of authenticating the transmitter, while an adversary not aware of $s_k$ has a knowledge gap $\delta$ which reduces its chances of performing the same task (device authentication).} 
    \label{fig:multiple_rounds}
\end{figure}

Let $\delta$ be the knowledge gap between the legitimate receiver and the adversary, and $p$ be the probability that the legitimate receiver correctly classifies the transmitters. We can express the probability $p_\mathcal{A}$ that the adversary classifies correctly the transmitters as $p_\mathcal{A} = p - \delta$. Given a sequence $w$ of consecutive observations (images generated from a number of \ac{IQ} samples), any receiver (including unauthorized ones) can make an educated guess to maximize detection accuracy. Indeed, both the receiver and the adversary can collect $w$ observations and then apply a majority voting decision on independent observations to maximize their chances of guessing the actual labels to be assigned to the images (transmitting sources)---under the assumption that the seed $s_k$ is not changed during $w$. The probability of getting $v$ successes in $w$ independent trials is given by the binomial distribution according to Eq.~\ref{eq:binomial}.
\begin{equation} 
    P(X = v) = {w \choose v} p^v (1-p)^{w-v}.
    \label{eq:binomial}
\end{equation}
The overall outcome is successful if the number of successful RFF classifications $v$ exceeds $\lceil \frac{w}{2} \rceil$. Therefore, the probability that majority voting is successful yields from Eq.~\ref{eq:maj_voting}.
\begin{equation} 
    P_{succ} = \sum_{v = \lceil \frac{w}{2} \rceil}^w {w \choose v} p^v (1-p)^{w-v}.
    \label{eq:maj_voting}
\end{equation}
We tested our approach via Monte-Carlo simulations considering $p=0.96$ as the probability of successful RFF by the legitimate receiver, as previously obtained and reported in Fig.~\ref{fig:radio_cnn}. Moreover, we consider an adversary with different knowledge gaps in the range $\delta = \{0.1, 0.2, 0.3, 0.4, 0.5\}$. Our results are shown in Fig.~\ref{fig:multiple_rounds}, where we highlight with circle markers the probability $P_{succ}$ of successful \ac{RFF} by the legitimate receiver, considering majority voting over a growing number of rounds (x-axis). We use different symbols for various values of the knowledge gap $\delta$. While increasing the number of rounds $w$ makes the outcome of the majority voting more decisive, we observe how $\delta$ gives a significant advantage to the legitimate receiver. The value $w=6$ allows the legitimate receiver to experience an overwhelming probability of correctly guessing the transmitter (greater than 0.99). However, an adversary with a knowledge gap $\delta = 0.3$, i.e., $p_\mathcal{A} = 0.66$, experiences the same probability of success while adopting the majority voting, i.e., $P_{succ} \approx p_\mathcal{A}$. 
Finally, note that the transmitter can maximize the probability of successful \ac{RFF} $P_{succ}$ of the legitimate receiver as opposed to the one of non-authorized receivers, e.g., updating the seed $s_k$ every $w=3$ rounds.

\section{Discussion}
\label{sec:discussion}
{\bf Main Findings.} Our work confirms that \ac{RFF} can be used to perform de-anonymization attacks at the physical layer against radio devices while proposing an effective and efficient solution to prevent it. The core idea of our proposed solution, \sol, relies on injecting noise during the modulation process. The injected noise level should be, at the same time, high enough to hide the fingerprint and low enough not to affect the quality of the communication link. Experimental results leveraging both wired and wireless links show that the impact of the noise on the quality of the received signal is less than 0.1~dB while achieving a complete disruption of the fingerprint. We first collected data from a wired link to estimate the real impact of the noise on the fingerprint, excluding, on purpose, additional distortion and noise phenomena typical of the wireless channel, such as multipath fading. The ability to identify and track a device is hindered by the injection of noise. This can be achieved by considering any noise pattern and a standard deviation of $\sigma \ge 0.02$. We confirmed our findings for wired and wireless links while training on noise-free samples and testing on any other noise level.
We also confirmed the robustness of \sol\ considering two adversary models as a function of the adversary's capabilities to collect information from the radio devices. We considered different training configurations spanning between noise-free and noisy samples while taking into account either a target device or a pool of devices. All the considered adversary models confirm that the injection of noise prevents the identification and tracking of the device. Finally, we proposed a new form of \ac{RFF}: {\em selective device fingerprinting}. A transmitter can control the ability of the receiver to authenticate it (at the physical layer) by resorting to a shared secret. We can thus achieve effective \ac{RFF} for a controlled subset of receiving entities, i.e., for many but not for all receivers eavesdropping on the radio spectrum.

{\bf Robustness to Channel Equalization Techniques.} An orthogonal approach that the adversary could use to improve the accuracy of transmitter identification is to use channel equalization techniques. These techniques have shown a remarkable capability to correct transmission errors when the noise is introduced by the wireless channel. Note that channel equalization techniques work by assuming that the statistical properties of the wireless channel are predictable and stable over short periods, and that they do not change too quickly~\cite{haykin2008_book}. 
In our context, the transmitter using \sol\ continuously switches the noise type and the standard deviation of the injected noise ($\sigma$). This behavior breaks, by definition, any assumptions about the predictability and stability of the noise added to the signal, making those techniques unreliable. Moreover, channel equalization techniques have been conceived to recover the symbols (and thus, the bits) transmitted by the receiver, in the presence of distortions (noise and multipath). They do not recover the original value of the \ac{IQ} sample before noise addition. Thus, the application of those techniques reshapes (by design) the radio fingerprint experienced by the receiver.

{\bf Limitations.} We consider a limited set of 10 transmitting devices, in accordance with other reference RFF works~\cite{alhazbi2023_acsac, lu2024_arxiv}. We only considered the \ac{BPSK} modulation scheme, which is consistently used in many wireless technologies, making our results immediately applicable. We did not apply our solution to other higher-order modulation schemes. Although schemes like QPSK and 64-QAM could be more sensitive to noise, such modulation schemes are well robust to a degradation of the SNR of only 0.1 dB (see Fig.~\ref{fig:radio_cnn}). We also limited the type of experiments in the wireless scenario by considering a static wireless scenario, while we modelled propagation over larger distances by considering various SNR levels and an attenuator of 30dB. These scenarios are the ones where RFF exhibits more reliable performance, so this methodology does not affect our findings. Finally, we considered only the image-based RFF solutions based on {\em ResNet-18} and \acp{AE}, based on relevant recent literature~\cite{oligeri2023tifs},~\cite{alhazbi2023_acsac}, and RFF approaches based on raw IQ samples~\cite{liu2023_wcl},~\cite{sun2022_rs},~\cite{lu2024_arxiv}. In our future work, we will extend our analysis with further classifiers. 

\section{Conclusion and Future Work}
\label{sec:conclusion}

We have presented {\em \sol}, a technique to prevent hostile \ac{RFF} by injecting random noise in the emitted \ac{RF} signal without affecting communication quality, thus mitigating physical layer attacks aimed at identifying and tracking radio devices. We have tested our solution against state-of-the-art classification techniques, different adversary models, and scenarios. 
We have proven that the injection of a minimal amount of noise in the transmitted signal does not affect the quality of the link but completely hides the fingerprint from the receiver. Moreover, we have discussed the use of {\em \sol} for selective radio fingerprint disclosure, enabling only a subset of legitimate receivers to perform \ac{RFF} while excluding unauthorized ones.
Future works include testing our solution against different classifiers while considering different scenarios.


\bibliographystyle{IEEEtran}
\balance
\bibliography{main}

@ARTICLE{bechir_2024_tmlcn,
  author={Elmaghbub, Abdurrahman and Hamdaoui, Bechir},
  journal={IEEE Transactions on Machine Learning in Communications and Networking}, 
  title={Distinguishable IQ Feature Representation for Domain-Adaptation Learning of WiFi Device Fingerprints}, 
  year={2024},
  volume={2},
  number={},
  pages={1404-1423},
  doi={10.1109/TMLCN.2024.3446743}}

@inproceedings{alhazbi2023_acsac,
  title={The {Day-After-Tomorrow}: On the performance of radio fingerprinting over time},
  author={Alhazbi, Saeif and Sciancalepore, Savio and Oligeri, Gabriele},
  booktitle={Proc. of the 39th Annual Computer Security Applications Conference},
  pages={439--450},
  year={2023}
}

@ARTICLE{h2023_twc,
  author={He, Jiashuo and Huang, Sai and Chang, Shuo and Wang, Fanggang and Shen, Ba-Zhong and Feng, Zhiyong},
  journal={IEEE Transactions on Wireless Communications}, 
  title={{Radio Frequency Fingerprint Identification With Hybrid Time-Varying Distortions}}, 
  year={2023},
  volume={22},
  number={10},
  pages={6724-6736},
  doi={10.1109/TWC.2023.3245070}}

@article{solenthaler2025_spacesec,
  title={{OrbID: Identifying Orbcomm Satellite RF Fingerprints}},
  author={Solenthaler, C{\'e}dric and Strohmeier, Joshua Smailes Martin},
  journal={3rd Workshop on the Security of Space and Satellite Systems},
  year={2025}
}

@inproceedings{elmaghbub2024_wisec,
author = {Elmaghbub, Abdurrahman and Hamdaoui, Bechir},
title = {{No Blind Spots: On the Resiliency of Device Fingerprints to Hardware Warm-Up Through Sequential Transfer Learning}},
year = {2024},
doi = {10.1145/3643833.3656138},
booktitle = {Proceedings of the 17th ACM Conference on Security and Privacy in Wireless and Mobile Networks},
pages = {134–144},
numpages = {11},
location = {Seoul, Republic of Korea},
series = {WiSec '24}
}

@inproceedings{givehchian2024_sp,
  title={{Practical obfuscation of BLE physical-layer fingerprints on mobile devices}},
  author={Givehchian, Hadi and Bhaskar, Nishant and Redding, Alexander and Zhao, Han and Schulman, Aaron and Bharadia, Dinesh},
  booktitle={2024 IEEE Symposium on Security and Privacy (SP)},
  pages={2867--2885},
  year={2024},
  organization={IEEE}
}

@ARTICLE{papangelo2023commag,
  author={Papangelo, Lorenzo and Pistilli, Maurizio and Sciancalepore, Savio and Oligeri, Gabriele and Piro, Giuseppe and Boggia, Gennaro},
  journal={IEEE Communications Magazine}, 
  title={{Adversarial Machine Learning for Image-Based Radio Frequency Fingerprinting: Attacks and Defenses}}, 
  year={2024},
  volume={},
  number={},
  pages={1-7},
  doi={10.1109/MCOM.001.2300464}}

@article{baig2025_tai,
  title={{Leveraging AI to Compromise IoT Device Privacy by Exploiting Hardware Imperfections}},
  author={Baig, Mirza Athar and Iqbal, Asif and Aman, Muhammad Naveed and Sikdar, Biplab},
  journal={IEEE Transactions on Artificial Intelligence},
  year={2025},
  publisher={IEEE}
}

@inproceedings{irfan2024_asiaccs,
author={Irfan, Muhammad and Sciancalepore, Savio and Oligeri, Gabriele},
title={Preventing Radio Fingerprinting through Low-Power Jamming},
booktitle={ACM Asia Conference on Computer and Communications Security (ASIA CCS ’25)}, 
year={2025},
doi={10.1145/3708821.3733862}}

@INPROCEEDINGS{alhazbi2023ccnc,
  author={Alhazbi, Saeif and Sciancalepore, Savio and Oligeri, Gabriele},
  booktitle={IEEE Consumer Communications \& Networking Conference (CCNC)}, 
  title={BloodHound: Early Detection and Identification of Jamming at the PHY-layer}, 
  year={2023},
  volume={},
  number={},
  pages={1033-1041},
  doi={10.1109/CCNC51644.2023.10059878}}

@ARTICLE{sciancalepore2023jamming,
  author={Sciancalepore, Savio and Kusters, Fabrice and Abdelhadi, Nada Khaled and Oligeri, Gabriele},
  journal={IEEE Internet of Things Journal}, 
  title={{Jamming Detection in Low-BER Mobile Indoor Scenarios via Deep Learning}}, 
  year={2024},
  volume={11},
  number={8},
  pages={14682-14697},
}

@INPROCEEDINGS{oligeri2024sac,
  author={Sadighian, Alireza and Sciancalepore, Savio and Oligeri, Gabriele},
  booktitle={2024 ACM Symposium on Applied Computing (SAC)}, 
  title={{SatPrint}: Satellite Link Fingerprinting}, 
  year={2024},
  volume={},
  number={},
  pages={177-185},
  doi={10.1145/3605098.3636011}}

@INPROCEEDINGS{sadighian2024ccnc,
  author={Sadighian, Alireza and Sciancalepore, Savio and Oligeri, Gabriele},
  booktitle={2024 IEEE 21th Consumer Communications \& Networking Conference (CCNC)}, 
  title={{FadePrint}: Satellite Spoofing Detection via Fading Fingerprinting}, 
  year={2024},
  volume={},
  number={},
  doi={10.1109/CCNC51644.2023.10059878}}

@ARTICLE{oligeri2023tifs,
  author={Oligeri, Gabriele and Sciancalepore, Savio and Raponi, Simone and Pietro, Roberto Di},
  journal={IEEE Transactions on Information Forensics and Security}, 
  title={{PAST-AI}: Physical-Layer Authentication of Satellite Transmitters via Deep Learning}, 
  year={2023},
  volume={18},
  number={},
  pages={274-289},
  doi={10.1109/TIFS.2022.3219287}}

@INPROCEEDINGS{alhazbi2024_iwcmc,
  author={Al-Hazbi, Saeif and Hussain, Ahmed and Sciancalepore, Savio and Oligeri, Gabriele and Papadimitratos, Panos},
  booktitle={International Wireless Communications and Mobile Computing (IWCMC)}, 
  title={{Radio Frequency Fingerprinting via Deep Learning: Challenges and Opportunities}}, 
  year={2024},
  volume={},
  number={},
  pages={0824-0829},
  doi={10.1109/IWCMC61514.2024.10592579}}

@inproceedings{deng2009imagenet,
  title={Imagenet: A large-scale hierarchical image database},
  author={Deng, Jia and Dong, Wei and Socher, Richard and Li, Li-Jia and Li, Kai and Fei-Fei, Li},
  booktitle={2009 IEEE conference on computer vision and pattern recognition},
  pages={248--255},
  year={2009},
  organization={Ieee}
}

@inproceedings{Samarati1998ProtectingPW,
  title={Protecting privacy when disclosing information: k-anonymity and its enforcement through generalization and suppression},
  author={Pierangela Samarati and Latanya Sweeney},
  year={1998},
}

@article{dipietro_2017_cn,
title = {Enabling broadcast communications in presence of jamming via probabilistic pairing},
journal = {Computer Networks},
volume = {116},
pages = {33-46},
year = {2017},
issn = {1389-1286},
author = {Roberto {Di Pietro} and Gabriele Oligeri},
}

@article{sun2022_rs,
  title={{Robustness of Deep Learning-Based Specific Emitter Identification under Adversarial Attacks}},
  OPTauthor={Sun, Liting and Ke, Da and Wang, Xiang and Huang, Zhitao and Huang, Kaizhu},
  author={{L. Sun, et al.}},
  journal={Remote Sensing},
  volume={14},
  number={19},
  pages={4996},
  year={2022},
  publisher={MDPI}
}

@inproceedings{zhuang2018_asiaccs,
  title={{Fbsleuth: Fake base station forensics via radio frequency fingerprinting}},
  author={Zhuang, Zhou and Ji, Xiaoyu and Zhang, Taimin and Zhang, Juchuan and Xu, Wenyuan and Li, Zhenhua and Liu, Yunhao},
  booktitle={Proceedings of the 2018 on Asia Conference on Computer and Communications Security},
  pages={261--272},
  year={2018}
}

@article{smailes2024_arxiv,
  title={{Sticky Fingers: Resilience of Satellite Fingerprinting against Jamming Attacks}},
  author={Smailes, Joshua and Salkield, Edd and K{\"o}hler, Sebastian and Birnbach, Simon and Strohmeier, Martin and Martinovic, Ivan},
  journal={arXiv preprint arXiv:2402.05042},
  year={2024}
}

@article{reising2015_tifs,
  title={{Authorized and rogue device discrimination using dimensionally reduced RF-DNA fingerprints}},
  author={Reising, Donald R and Temple, Michael A and Jackson, Julie A},
  journal={IEEE Transactions on Information Forensics and Security},
  volume={10},
  number={6},
  pages={1180--1192},
  year={2015},
  publisher={IEEE}
}

@article{reising2020_iotj,
  title={{Radio identity verification-based IoT security using RF-DNA fingerprints and SVM}},
  author={Reising, Donald and Cancelleri, Joseph and Loveless, T Daniel and Kandah, Farah and Skjellum, Anthony},
  journal={IEEE Internet of Things Journal},
  volume={8},
  number={10},
  pages={8356--8371},
  year={2020},
  publisher={IEEE}
}

@inproceedings{smailes2023_ccs,
  title={{Watch This Space: Securing Satellite Communication through
Resilient Transmitter Fingerprinting}},
  author={Smailes, Joshua and K{\"o}hler, Sebastian and Birnbach, Simon and Strohmeier, Martin and Martinovic, Ivan},
  booktitle={Proceedings of the 2023 ACM SIGSAC Conference on Computer and Communications Security},
  pages={608--621},
  year={2023}
}

@article{danev2012_csur,
  title={On physical-layer identification of wireless devices},
  author={Danev, Boris and Zanetti, Davide and Capkun, Srdjan},
  journal={ACM Computing Surveys (CSUR)},
  volume={45},
  number={1},
  pages={1--29},
  year={2012},
  publisher={ACM New York, NY, USA}
}

@article{gu2024_tosn,
  title={{RF-TESI: Radio Frequency Fingerprint-based Smartphone Identification under Temperature Variation}},
  author={Gu, Xiaolin and Wu, Wenjia and Song, Aibo and Yang, Ming and Ling, Zhen and Luo, Junzhou},
  journal={ACM Transactions on Sensor Networks},
  volume={20},
  number={2},
  pages={1--21},
  year={2024},
  publisher={ACM New York, NY}
}

@article{shen2022_tifs,
  title={{Towards scalable and channel-robust radio frequency fingerprint identification for LoRa}},
  author={Shen, Guanxiong and Zhang, Junqing and Marshall, Alan and Cavallaro, Joseph R},
  journal={IEEE Transactions on Information Forensics and Security},
  volume={17},
  pages={774--787},
  year={2022},
  publisher={IEEE}
}

@inproceedings{restuccia2019_manet,
author = {Restuccia, Francesco and D'Oro, Salvatore and Al-Shawabka, Amani and Belgiovine, Mauro and Angioloni, Luca and Ioannidis, Stratis and Chowdhury, Kaushik and Melodia, Tommaso},
title = {{DeepRadioID: Real-Time Channel-Resilient Optimization of Deep Learning-based Radio Fingerprinting Algorithms}},
year = {2019},
isbn = {9781450367646},
publisher = {Association for Computing Machinery},
booktitle = {Proceedings of the Twentieth ACM International Symposium on Mobile Ad Hoc Networking and Computing},
pages = {51–60},
numpages = {10},
location = {Catania, Italy},
series = {Mobihoc '19}
}

@ARTICLE{xu2016_comst,
  author={Xu, Qiang and Zheng, Rong and Saad, Walid and Han, Zhu},
  journal={IEEE Communications Surveys \& Tutorials}, 
  title={{Device Fingerprinting in Wireless Networks: Challenges and Opportunities}}, 
  year={2016},
  volume={18},
  number={1},
  pages={94-104},
  }

@inproceedings{danev2010_wisec,
author = {Danev, Boris and Luecken, Heinrich and Capkun, Srdjan and El Defrawy, Karim},
title = {{Attacks on physical-layer identification}},
year = {2010},
isbn = {9781605589237},
booktitle = {Proceedings of the Third ACM Conference on Wireless Network Security},
pages = {89–98},
numpages = {10},
location = {Hoboken, New Jersey, USA},
series = {WiSec '10}
}

@book{haykin2008_book,
  title={{Communication Systems}},
  author={Haykin, Simon},
  year={2008},
  publisher={John Wiley \& Sons}
}

@inproceedings{alshawabka2020_infocom,
  title={{Exposing the fingerprint: Dissecting the impact of the wireless channel on radio fingerprinting}},
  author={Al-Shawabka, Amani and Restuccia, Francesco and D’Oro, Salvatore and Jian, Tong and Rendon, Bruno Costa and Soltani, Nasim and Dy, Jennifer and Ioannidis, Stratis and Chowdhury, Kaushik and Melodia, Tommaso},
  booktitle={IEEE INFOCOM 2020-IEEE Conference on Computer Communications},
  pages={646--655},
  year={2020},
  organization={IEEE}
}

@inproceedings{irfan2024_arxiv_reliability,
  title={{Shifting Signatures: The Ephemeral Nature of the Radio Fingerprint on the USRP X310}},
  author={Irfan, Muhammad and Sciancalepore, Savio and Oligeri, Gabriele},
  booktitle={IEEE International Conference on Communications and Network Security (CNS)},
  year={2025},
  organization={IEEE}
}

@book{rappaport2024_book,
  title={{Wireless communications: principles and practice}},
  author={Rappaport, Theodore S},
  year={2024},
  publisher={Cambridge University Press}
}

@TECHREPORT{X410_datasheet,
 author = {{National Instruments}},
 title = {{NI USRP X410 Specification
}},
 type = {Data Sheet}
}

@misc{dataset_anon,
  title = {{Open Source Data}},
  author={Anonymous},
  howpublished = {},
  note = {Accessed: 2025-Aug-19}
}

@TECHREPORT{b200_datasheet,
 author = {{National Instruments}},
 title = {{NI USRP b200 mini-i Specification}},
 type = {Data Sheet}
}

@misc{sampling_same,
	author = {{DSP StackExchange}},
	title = {{Bandwidth with complex sampling}},
	howpublished = {\url{https://dsp.stackexchange.com/questions/36927/bandwidth-with-complex-sampling}},
	note = {(Accessed: 2025-Aug-19)},
	month={Apr.},
	year = {2017}
}

@misc{lu2024_arxiv,
      title={{Erasing Radio Frequency Fingerprints via Active Adversarial Perturbation}}, 
      author={Zhaoyi Lu and Wenchao Xu and Ming Tu and Xin Xie and Cunqing Hua and Nan Cheng},
      year={2024},
      eprint={2406.07349},
      archivePrefix={arXiv},
      primaryClass={cs.CR},
      url={https://arxiv.org/abs/2406.07349}, 
}

@article{jagannath2022_comnet,
  title={{A comprehensive survey on radio frequency ({RF}) fingerprinting: Traditional approaches, deep learning, and open challenges}},
  author={Jagannath, Anu and Jagannath, Jithin and Kumar, Prem Sagar Pattanshetty Vasanth},
  journal={Computer Networks},
  volume={219},
  pages={109455},
  year={2022},
  publisher={Elsevier}
}

@inproceedings{ardoin2025_arxiv,
  title={Tracking UWB devices through radio frequency fingerprinting is possible},
  author={Ardoin, Thibaud and Pauli, Niklas and Gro{\ss}, Benedikt and Kholghi, Mahsa and Reaz, Khan and Wunder, Gerhard},
  booktitle={2025 International Conference on Computing, Networking and Communications (ICNC)},
  pages={394--400},
  year={2025},
  organization={IEEE}
}

@article{abanto2020_macs,
  title={{Stay Connected, Leave no Trace: Enhancing Security and Privacy in WiFi via Obfuscating Radiometric Fingerprints}},
  author={Abanto-Leon, Luis Fernando and B{\"a}uml, Andreas and Sim, Gek Hong and Hollick, Matthias and Asadi, Arash},
  journal={Proceedings of the ACM on Measurement and Analysis of Computing Systems},
  volume={4},
  number={3},
  pages={1--31},
  year={2020},
  publisher={ACM New York, NY, USA}
}

@article{liu2023_wcl,
  title={{Robust Adversarial Attacks on Deep Learning Based RF Fingerprint Identification}},
  OPTauthor={Liu, Boyang and Zhang, Haoran and Wan, Yiyao and Zhou, Fuhui and Wu, Qihui and Ng, Derrick Wing Kwan},
  author={{B. Liu, et al.}},
  journal={IEEE Wireless Communications Letters},
  year={2023},
  publisher={IEEE}
}

\end{document}